\documentclass[useAMS]{mn2e}
\usepackage{lscape}
\usepackage{rotating}
\usepackage{amssymb}

\def\msun{\hbox{M$_\odot$}}
\def\mstar{\hbox{M$_{\rm c}$}}
\def\t4{\hbox{t$_{\rm 4}$}}
\def\mc{\hbox{M$_{\rm c}$}}

\def\fmid{\hbox{F$_{\rm MID}$}}
\def\reff{\hbox{R$_{\rm eff}$}}

\newcommand{\cfr}{{\rm CFR}}
\newcommand{\dr}{{\rm d}}
\newcommand{\dlg}{\Delta_{\lambda\gamma}}
\newcommand{\dndt}{\frac{\dr N}{\dr t}}

\newcommand{\dndm}{\frac{\dr N}{\dr M}}
\newcommand{\dndmdt}{\frac{\dr^2 N}{\dr M\dr t}}
\newcommand{\dndmidt}{\frac{\dr^2 N}{\dr\mi\dr t}}
\newcommand{\dndmidttext}{\dr N/(\dr\mi\dr t)}
\newcommand{\dmdm}{\frac{\partial\mi}{\partial M}}
\newcommand{\eone}{{\rm E}_{1}}
\newcommand{\en}{{\rm E}_{n}}
\newcommand{\funcmid}{F_{\rm MID}}
\newcommand{\mdot}{\dot{M}}
\newcommand{\mdotd}{\dot{M}_{\rm MDD}}
\newcommand{\mdoti}{\dot{M}_{\rm MID}}
\newcommand{\mi}{M_i}
\newcommand{\mlo}{M_{\rm lo}}
\newcommand{\mup}{M_{\rm up}}
\newcommand{\sfr}{{\rm SFR}}
\newcommand{\tdis}{t_{\rm dis}}
\newcommand{\tn}{t_0}
\newcommand{\tmin}{t_{\rm min}}
\newcommand{\tmax}{t_{\rm max}}
\newcommand{\ts}{t_{\rm s}}

\input psfig.sty
\input epsf.sty
\usepackage{graphicx}
\usepackage{epsfig}
\title[M83 Stellar Cluster Population]
{Stellar Clusters in M83: Formation, evolution, disruption and the influence of environment}
\author[Bastian et al.] {N. Bastian$^{1}$, A. Adamo$^{2}$, M. Gieles$^3$, E. Silva-Villa$^4$, H.J.G.L.M Lamers$^4$, S.S. Larsen$^4$,  \newauthor L.J. Smith$^5$, I.S. Konstantopoulos$^6$,  \& E. Zackrisson$^2$ \\
$^1$ Excellence Cluster Universe, Boltzmannstr. 2, 85748 Garching, Germany\\
$^2$ Department of Astronomy, Stockholm University, Oscar Klein Centre, AlbaNova, Stockholm SE-106 91, Sweden\\
$^3$ Institute of Astronomy, University of Cambridge, Madingley Road, Cambridge CB3 0HA, UK\\
$^4$ Astronomical Institute, Utrecht University, Princetonplein 5, NL-3584CC Utrecht, the Netherlands\\
$^5$ Space Telescope Science Institute and European Space Agency, 3700 San Martin Drive, Baltimore, MD 21218, USA\\
$^6$ Department of Astronomy and Astrophysics, The Pennsylvania State University, University Park, PA 16802, USA\\
}
\date{Accepted. Received; in original form}
\pagerange{\pageref{firstpage}--\pageref{lastpage}}
\pubyear{2011}
\begin{document}
\maketitle
\label{firstpage}
\begin{abstract}
We study the stellar cluster population in two adjacent fields in the nearby, face-on spiral galaxy, M83, using multi-wavelength WFC3/HST imaging.  After automatic detection procedures, the clusters are selected through visual inspection to be centrally concentrated, symmetric, and resolved on the images, which allows us to differentiate between clusters and likely unbound associations.  We compare our sample with previous studies and show that the differences between the catalogues are largely due to the inclusion of large numbers of diffuse associations within previous catalogues as well as the inclusion of the central starburst region, where the completeness limit is significantly worse than in the surrounding regions.  We derive the size distribution of the clusters, which is well described by a log-normal distribution with a peak at $\sim2.5$~pc, and find evidence for an expansion in the half-light radius of clusters with age.  The luminosity function of the clusters is well approximated by a power-law with index, $-2$, over most of the observed range, however a steepening is seen at M$_{\rm V} = -9.3$ and $-8.8$ in the inner and outer fields, respectively. Additionally, we show that the cluster population is inconsistent with a pure power-law mass distribution, but instead exhibits a truncation at the high mass end.   If described as a Schechter function, the characteristic mass is 1.6 and 0.5 $\times10^5$~\msun, for the inner and outer fields, respectively, in agreement with previous estimates of other cluster populations in spiral galaxies.  Comparing the predictions of the  {\it mass independent disruption} (MID)  and {\it mass dependent disruption} (MDD) scenarios with the observed distributions, we find that both models can accurately fit the data.  However, for the MID case, the fraction of clusters destroyed (or mass lost) per decade in age is dependent on the environment, hence, the age/mass distributions of clusters are not universal.  In the MDD case, the disruption timescale scales with galactocentric distance (being longer in the outer regions of the galaxy) in agreement with analytic and numerical predictions.  Finally, we discuss the implications of our results on other extragalactic surveys, focussing on the fraction of stars that form in clusters and the need (or lack thereof) for infant mortality.

\end{abstract}

\section{Introduction}
\label{sec:intro}

A by-product of the star formation process is the formation of stellar groups, some of which are gravitationally bound and likely long-lived, while other groups are unbound and will disperse into the field.  In the Galaxy, the former are often referred to as open clusters while the latter are known as associations.  However, for most extragalactic studies the limited spatial resolution, even when using {\it Hubble Space Telescope} imaging, as well as the lack of dynamical information for individual stars, makes the distinction between these two groups challenging.  This is particularly difficult at young ($<10$~Myr) ages before significant dynamical evolution has taken place.  Such ambiguities can severely affect the interpretation of extragalactic cluster samples, especially when large numbers of young groups are present, as in starburst galaxies.  For galaxies that have a dominant older population, such as post-merger remnants like NGC~7252 (Schweizer \& Seitzer 1998) or NGC 1316 (Goudfrooij et al.~2001), clusters can be straightforwardly defined due to their compact nature and the fact that associations have dissolved into the field.

Using a sample of resolved stellar groups of various ages (including groups fully resolved into their constituent stars as well as groups that appear extended on images) compiled from the literature, Portegies Zwart, McMillan \& Gieles (2010, hereafter PZMG10) have shown that there exists a continuous distribution of stellar structures at young ages (from loose groups to dense clusters), while at older ages ($>10$~Myr) a bimodal distribution develops, with bound and unbound groups that are clearly distinguishable. This observation fits well with the inferred hierarchical distribution of star-formation, where no preferred scale exists from tenths to hundreds of parsecs, meaning that clusters are not distinct or unique objects at young ages (e.g., Bastian~2011).  Hence, clusters cannot be defined until a stellar group is dynamically evolved.

%A review of the observed initial conditions and early evolution of stellar groups is given in Bastian~(2011).

Along similar lines, PZMG10 also showed that the derived properties of a cluster population depend on how clusters are defined.  For example, the age distribution of loose stellar structures (effective radii, \reff, greater than $6$~pc) in the SMC shows a rapid decline, with the number of groups dropping as a function of age as $t^{-1}$.  On the other hand, if only compact groups are considered (\reff\ $< 6$~pc) the age distribution is mostly flat, negating the need of large amounts of disruption.

%{\bf Schweizer 2004 paper talking about clusters/complexes}

In the present work, we study the cluster population of the nearby face-on spiral, M83, using new {\it Wide Field Camera 3} (WFC3) imaging with {\it HST}.  In order to define our cluster sample, each cluster candidate is assessed by eye, and only resolved and centrally concentrated objects are included in the sample.  The sample covers a large range in galactocentric distances, hence sampling a large range of environmental conditions.  Using our conservatively defined cluster sample, we have previously found evidence for a strong dependence of cluster lifetimes on the ambient environment (Bastian et al. 2011, hereafter B11).  Here, we use the same sample to study 1) the distribution of cluster sizes, luminosities, and masses, 2) empirical cluster disruption laws,  and 3) relations among the derived parameters (age, mass, radius). 

\subsection{Cluster Formation and Disruption}
\label{sec:intro-disruption}

The standard paradigm that has developed over recent years is that all or most stars form in clusters, and that a large fraction ($\sim90$\%) of the clusters are destroyed during the process of going from the embedded to the exposed phase (Lada \& Lada~2003)\footnote{However, recent studies on the spatial distribution of both high mass (e.g., Lamb et al.~2010; Bressert et al.~2011) and low mass (e.g., Bressert et al.~2010) young stars have shown that clusters may be the fundamental unit of star-formation.  Indeed, clusters are not expected to be if star-formation follows a hierarchical distribution which would result in stars forming in a continues distribution, from dense clusters to (near) isolation (c.f., Bastian~2011).}.  This process, known as infant mortality, is thought to be due to the removal of gas left over from the star-formation process.  This gas expulsion is expected to be largely independent of the ambient environment since gas expulsion is largely an internal process and independent of mass since the mass loss is driven by violent relaxation (e.g. Goodwin \& Bastian~2006).  This phase of a cluster's evolution is expected to last a few crossing times, namely $\lesssim 10$~Myr for massive dense clusters (PZMG10).

After this initial phase, the surviving clusters are not expected to live indefinitely, but rather to disrupt due to internal (e.g., two-body relaxation, stellar evolution) and external (e.g., tidal fields) processes.  Fall, Chandar \& Whitmore~(2009) have suggested that these processes happen over different timescales, and hence can be treated independently, in addition to being largely independent of cluster mass.  Alternatively, through an analytic approach, Gieles et al.~(2011) find that all processes are acting concurrently and that the lifetime of a cluster depends on the ambient environment and its mass, in agreement with other theoretical investigations (e.g., H\'enon 1961; Spitzer 1987).

Observationally, the situation is less clear, and the community has yet to reach a consensus regarding the amount of disruption observed in populations as well as the role of cluster mass and environment in the process.  There have been two main empirical disruption laws put forward in the literature, based on different cluster samples in a variety of galaxies (explained in detail below).  In only one case, the SMC, have the advocates for the different scenarios used the same dataset (Chandar, Fall \& Whitmore~2006; Gieles, Portegies Zwart \& Lamers~2007), and even there the authors come to different conclusions.  

The first empirical disruption law considered here is {\it Mass Dependent Disruption} (MDD - Lamers et al.~2005).  This was first presented in Boutloukos \& Lamers~(2003), assuming instantaneous disruption, in order to explain the observed cluster population properties in the Galaxy, SMC, M33 and M51 (however, see Larsen~(2008) for a thorough review of empirical mass dependent disruption laws).  In this scenario, the lifetime of a cluster is dependent on the initial mass of the cluster as $M^{\gamma}$ with $\gamma \sim 0.62$, and on the ambient environment with clusters surviving longer in galaxies with weak tidal forces and low numbers of GMCs.  The empirical model was updated in Lamers et al.~(2005; this is the form that we adopt here, which includes gradual cluster mass loss), applied to a distance limited sample of open clusters in the Milky Way in Lamers \& Gieles~(2006), and was shown to agree with predictions from numerical $N$-body experiments in Gieles et al.~(2004).

The second empirical disruption law considered here is {\it Mass Independent Disruption} (MID - Whitmore et al.~2007).  In this scenario cluster disruption is independent on cluster mass and the ambient environment.  While the classic infant mortality falls in this category and is thought to last for $\lesssim10$~Myr (e.g. Lada \& Lada~2003), the concept has been expanded up to ages of $\sim1$~Gyr and has been invoked to explain cluster populations in the Antennae (Fall et al.~(2005), the LMC (Chandar et al. 2010a), and the inner regions of M83 (Chandar et al. 2010b, hereafter C10, Fouesneau et al.~2011).  The model is described in detail in Whitmore et al.~(2007). This disruption scenario results in ``universal" age and mass distributions, with the number of clusters $d^2N_{\rm clusters}/dMdt \propto t^{-1}M^{-2}$, as cluster disruption dominates (over formation) the shape of the distributions.

Elmegreen \& Hunter~(2010) and Kruijssen et al.~(2011a) have shown that MID can be an emergent result of MDD.  In this scenario clusters are destroyed through strong interactions with the hierarchical interstellar medium, a process that may be (largely) mass independent if the shocks are strong enough but should depend strongly on the ambient density of the gas (i.e. environment). 

One of the goals of the current study is to distinguish between these two models, and place additional constraints upon them.  Additionally, we would like to understand the differences between the age and mass distributions derived in the different studies.  The new M83 WFC3 images provide a rich dataset with which we can test each of the possibilities for the differences reported in the literature.  In B11, we investigated possible differences between the ages/masses derived for the clusters in the dataset used here and the catalogue used in C10 and Whitmore et al.~(2011, hereafter W11).  We found a good agreement in the derived ages/masses between the two datasets, when a source appeared in both catalogues.  We concluded that the likely differences derived between the respective groups was due to sample selection and/or differing analysis methods.  

In B11, we used the datasets presented here to study the dependence of environment on the cluster disruption process.  Comparing the distribution of clusters in colour-colour space as well as through quantitative comparison between the age/mass distributions between the two fields, we found clear evidence for the strong influence of environment on cluster disruption.  The inner region of the galaxy was found to have a relatively high rate of cluster disruption, in agreement with C10, while in the outer regions clusters appeared to be much longer lived.  This is in agreement with predictions of the MDD scenario, and was attributed to the different GMC densities and tidal fields experienced by the clusters in the two fields.

\subsection{Cluster size distribution}

In spite of the nearly six orders of magnitude in observed cluster masses, the radii of clusters appear to occupy a relatively narrow range, with effective radii largely contained between 0.5 and 10~pc in effective (half-light) radii (e.g. PZMG10).  The large range of mass and narrow range of size means that any cluster mass/radius relation must be inherently shallow, which is unexpected given the strong mass/radius relation of star-forming molecular clouds and cores (c.f. Ashman \& Zepf~2001). The observed radius distribution of clusters is in general adequately described by a log-normal distribution with a peak at 3-4~pc.  This has been found for young populations in M51 (Scheepmaker et al.~2007) and M101 (Barmby et al.~2006) as well as for populations of old globular clusters (e.g. Jord\'an et al.~2005).  In fact, the peak of the distribution has been suggested to be a distance indicator (Jord\'an et al.~2005).

Here we investigate whether the size distribution of clusters in M83 is the same as found for other young cluster systems as well as any dependence of the size on the age or mass of the cluster.  This will provide constraints on the formation and evolution of clusters.  Additionally, along with the mass of a cluster, we will use the size to estimate the dynamical stability of stellar groups (Gieles \& Portegies Zwart~2011; - hereafter GPZ11) in order to see how well our cluster selection criteria work.

This paper is organised as follows.  In \S~\ref{sec:obs} we introduce the data used, the sample selection, photometry, and fitting techniques, and compare our sample to previous studies. In \S~\ref{sec:results} we present the results of the study.  We constrain cluster disruption models in \S~\ref{sec:disruption} and in \S~\ref{sec:discussion} we discuss some of the implications of our results.  Finally, we summarise our main findings in \S~\ref{sec:conclusions}.

\section{Observations}
\label{sec:obs}

The data used in the present work were partially presented in C10 (who studied just the inner field) as well as in B11 (who studied both fields), and are briefly summarised here.   The imaging consists of Early Release Science data (GO 11360, PI O'Connell) taken with the {\it UVIS} and {\it IR} detectors of the {\it WFC3} on HST of two adjacent fields in M83. The ``inner field" covers the nucleus of the galaxy as well as the inner spiral arms, while the ``outer field" extends the contiguous coverage into the outer parts of the galaxy.  For each pointing, images in the F336W (U - 2225s), F438W (B - 1840s), F657N (H$\alpha$ - 1484s), F814W (I - 1213s) were taken with {\it UVIS-WFC3}.  The numbers inside the parenthesis are the exposure times in each of the filters.  Additionally, the dataset also contains V-band images, although the specific filters differ between the two pointings, F555W (1203s) and the F550M (1203s) filters were used in the inner and outer fields, respectively.  Both of these filters will be referred to as ``V-band", although no transformations between any filters and the corresponding Cousins-Johnson filter system were performed.  Additionally, we also use two pointings of the  {\it IR-WFC3} detector with the F110W (J - 1797s) and F160W (H - 2397s) filters.  These images do not cover the full field of view of the optical WFC3 images, as the {\it IR} channel has a slightly smaller field of view.   All data was taken from the WFC3 Early Release Science website\footnote{http://archive.stsci.edu/prepds/wfc3ers/m83datalist.html} fully reduced. 

\begin{figure}
\centering
\includegraphics[width=8cm]{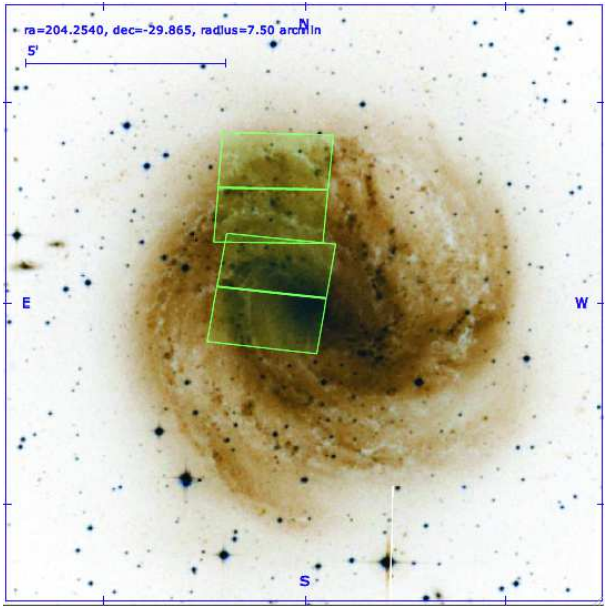}
\caption{A DSS image of M83 along with the WFC3-UVIS FOV for each field superimposed.}
\label{fig:image_full}
\end{figure}

\begin{figure}
\centering
\includegraphics[width=8cm]{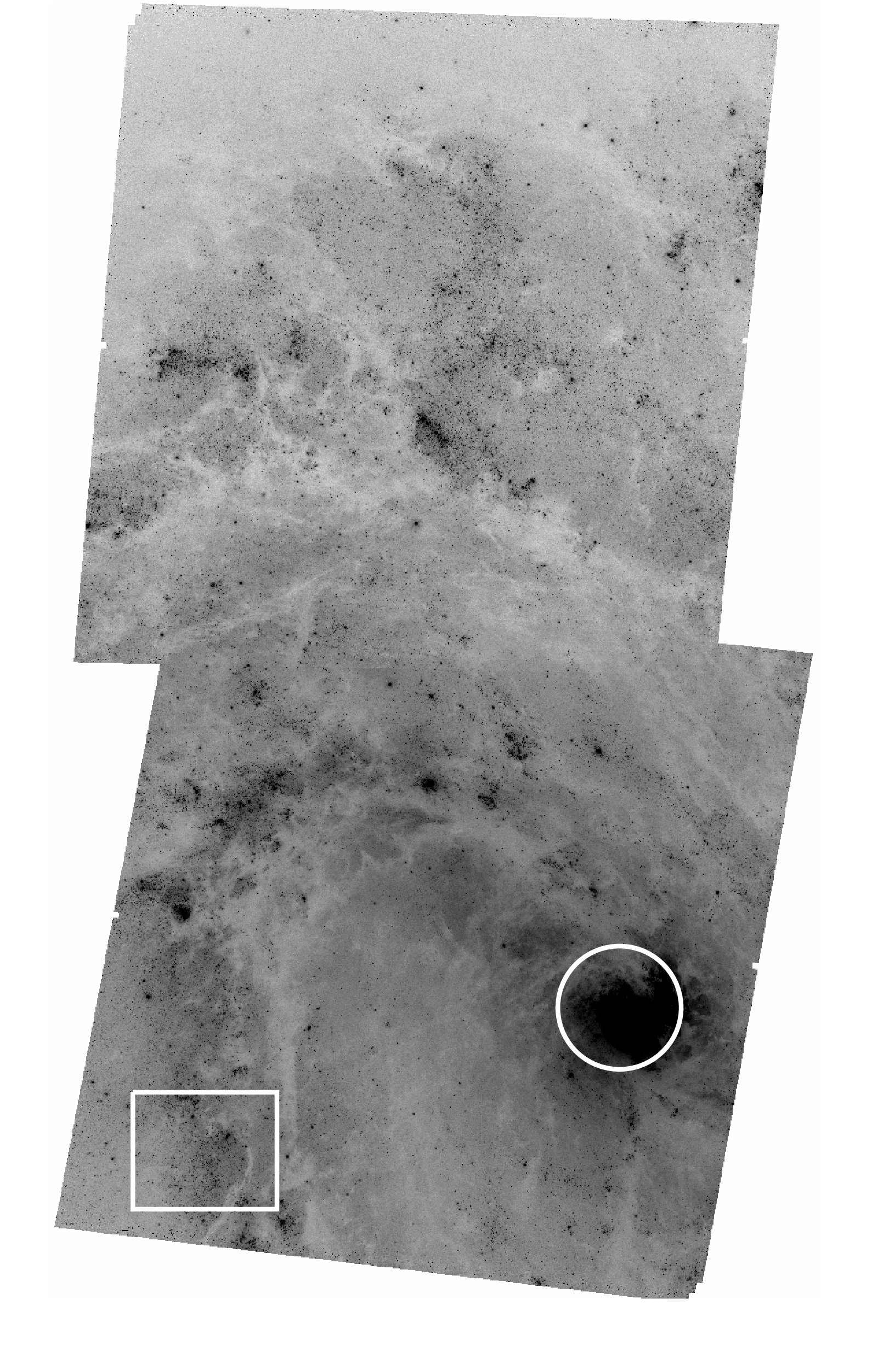}
\caption{The F438W (B) band image (logarithmically scaled) used in the present work.  The circle denotes the inner $\sim450$~pc that is not included in the present work, while the region indicated by the box is shown in more detail in Fig.~\ref{fig:region}.}
\label{fig:image_wfc3}
\end{figure}

The positions of the WFC3 pointings are shown in Fig.~\ref{fig:image_full} and the F438W (B) band image of the field of view is shown in Fig.~\ref{fig:image_wfc3}.  As in C10 and B11, we adopt a distance to M83 of $4.5$~Mpc (Thim et al.~2003).

\subsection{Sample Selection}
\label{sec:sample}

It is often assumed that all stars form in ``clusters", although defining a ``young cluster", as opposed to an association, is non-trivial.  Even within the solar neighbourhood, the estimated fraction of young stellar objects (YSOs) in clusters varies from $\sim20$\% to $\sim90$\% depending on how clusters are defined (Bressert et al.~2010). This confusion is likely the result of the hierarchical distribution of star-formation (e.g. Efremov \& Elmegreen~1998; Elmegreen~2002) where no distinct borders exist between clusters (or even subclusters), associations and complexes.  In this view, clusters are merely the densest structures in a continuous distribution of star-formation, and do not represent any unique spatial scale (as no unique spatial scale exists within an hierarchical or fractal distribution).   The importance of cluster selection, and the caveats associated with different techniques is discussed in detail in Silva-Villa \& Larsen (2011).

GPZ11 have suggested a dynamical definition of clusters, namely that the age of the stellar group (the stellar age, assuming a single formation epoch) is greater than the instantaneous crossing time (based on the estimated mass and radius) and define a quantity, $\Pi = t_{\rm age}/t_{\rm cross}$.  Unbound associations (on any scale) will expand with time, causing the crossing time to increase in proportion with age, meaning the $\Pi$ will stay at or below unity.  On the other hand, clusters will remain compact (although there may be some expansion - see Bastian et al.~2008 and \S~\ref{sec:size}) meaning that $\Pi$ will increase with age.  By applying this criterion to a sample of young stellar groups in the Galaxy, local group and nearby galaxies, GPZ11 found that for groups with ages less than 10~Myr, $\Pi$ was a continuous distribution, with no distinct break at $\Pi = 1$.  However, for groups older than 10~Myr, the $\Pi$ distribution became bi-modal, with clusters ($\Pi > 1$) and associations ($\Pi < 1$) becoming distinct populations.

For observational studies of extragalactic systems, where resolution effects limit the ability to study the structure of regions in detail, the result is that it is impossible to define a complete cluster sample at young ages ($<10$~Myr).  After this age, compact and centrally concentrated stellar structures are likely to be clusters, and it is possible to compile a well defined sample.  The implications of this for extragalactic surveys are discussed in more detail in \S~\ref{sec:discussion}.

With these caveats in mind, below we discuss how our cluster sample was defined.  As a first step we ran {\em SExtractor} (Bertin \& Arnouts 1996) over the B, V and I images with settings chosen to select a large number of candidates, including many false detections. These three catalogues were then cross-correlated, and only candidates detected in all three images were kept.  We then used {\em ISHAPE} (Larsen 1999) to estimate the size of each of the groups, and only selected resolved objects (FWHM $>$ 0.2 pixels; this is the limit where resolved/unresolved sources can be reliably distinguished). The PSF used for {\em ISHAPE} was derived for each filter independently, using bright, isolated, unresolved sources (i.e. stars) on the images. The V-band image of each cluster candidate was then examined, and only resolved, centrally concentrated and symmetric sources were retained as clusters. Approximately 40-50\% of the initial candidates were removed during this step.  A sample of objects that were excluded at this stage is labeled in Fig.~\ref{fig:region} and shown in detail in Fig.~\ref{fig:example_chandar}.  It is likely that some fraction of the stars in the groups presented in Fig.~\ref{fig:example_chandar} will end up in clusters after the system has dynamically evolved (e.g. the individual peaks in c13861), but it is not possible to tell what fraction based on imaging alone.  

We have adopted relatively strict selection criteria for selecting our sample in order to be able to compare our observations with models of cluster evolution and disruption, which assume that clusters are dense centrally concentrated objects.  We emphasise that visual inspection of cluster candidates necessarily introduces a level of subjectivity in the process, hence the catalogues of cluster candidates may differ between different authors.  This point must be kept in mind when interpreting the results and will be discussed in detail in \S~\ref{sec:comparison}.  We note that this two step process of visual inspection of the source candidates found by automated methods, has been found necessary in other recent works (e.g. Larsen~2004; Silva-Villa \& Larsen 2010, 2011; Annibali et al.~2011).

Importantly, we note that age was {\it not} used in the cluster selection procedure.  As will be discussed below, young stellar groups were preferentially excluded from our sample due to the application of the above criteria.  The implications of this will be discussed in more detail in \S~\ref{sec:discussion}.

Finally, we note that the selection of clusters using visual inspection is affected by the resolution of the observations.  In relatively distant systems, like the Antennae galaxies (e.g., Whitmore et al.~2010) or NGC~3256 (e.g., Goddard, Bastian, \& Kennicutt~2010), it will be more difficult to distinguish between clusters and associations, so these studies likely contain a larger fraction of associations in their cluster catalogues.  This point will be further discussed in \S~\ref{sec:discussion}.

\subsection{Photometry and Fitting}
\label{sec:photometry}

Once we have a produced a cluster catalogue, we carried out aperture photometry for each of the sources.  We adopted a aperture size for all bands of 5 pixels, and a background annulus with inner and outer radii of 8 and 10 pixels for the WFC3 UVIS images, respectively.  For the IR channel, we used the same physical size apertures and background, i.e., accounting for the differing pixel sizes.  The zeropoints were taking from the 
STScI webpage\footnote{http://www.stsci.edu/hst/wfc3/phot\_zp\_lbn}.  
Additionally, we corrected the F336W zeropoint for an $\sim4$\% additional efficiency (priv. comm. Jason Kalirai).  Additionally, we carried out photometry of our catalogue sources (only for the inner field) adopting the same parameters as Chandar et al.~(2010) and compared the results with their published catalogue.  Comparison between our photometry and that of Whitmore et al. (2011, hereafter W11) shows good agreement (mean deviations in colour $<0.03$~mag, accounting for the F336W shift, which was not applied in W11).  We note that the correction of the 4\% additional efficiency does not significantly affect the derived properties of the clusters.

Aperture corrections for each of the filters were derived from a sample of $\sim15$ resolved clusters on the images.  The resulting corrections from 5 to 15 pixels are: 0.43, 0.41, 0.41, 0.43, 0.44 \& 0.45~mag for the F336W, F438W, F547M, F555W, F657N \& F814W, respectively.  We added an additional 0.05~mag error in quadrature to the measured error for aperture correction uncertainties.   Finally, we corrected for Galactic extinction (Schlegel et al.~1998).

Once we have performed aperture photometry on the cluster catalogue, we proceed to estimate the age, mass and extinction of each of the candidates.  Only clusters with photometry in U, B, V and I are included.  The properties were estimated by comparing Simple Stellar Population (SSP) models to the observations.  We adopted the fitting procedure of Adamo et al. (2010a,b) and the Yggdrasil SSP models
(Zackrisson et al. 2011) for a Kroupa (2001) IMF, using Starburst99 Padova-AGB stellar population spectra (Leitherer et al. 1999, Vazquez \& Leitherer 2005) for $\sim2.5$ times solar metallicity ($z=0.050$; Bresolin \& Kennicutt 2002).  The fits were carried out including the U, B, V, I and H$\alpha$ photometry.  The Yggdrasil SSP models include nebular and continuum emission, allowing the H$\alpha$ fluxes to be used to differentiate young extincted clusters from older non-extincted clusters that have similar $U-B$ and $V-I$ colours.  The impact of including (or not) the J and H-band photometry will be discussed in detail in a forthcoming paper (Adamo et al., in prep.).

%We adopted the fitting procedure of Adamo et al. (2010a,b) and the Yggdrasil SSP models (Zackrisson et al.~2011) for a Kroupa~(2001) IMF and twice solar metallicity (Bresolin \& Kennicutt~2002).  

The masses derived in this way are the current masses, not the initial masses, meaning that stellar evolutionary mass loss has been taken into account (i.e. the initial masses are 10-30\% higher, depending on the age, if there would have been no disruption).  

The catalogue of the measured (magnitudes, positions, sizes) and estimated (age, mass, extinction) properties for the sample is given on the CDS website.

\begin{figure}
\includegraphics[width=8.5cm]{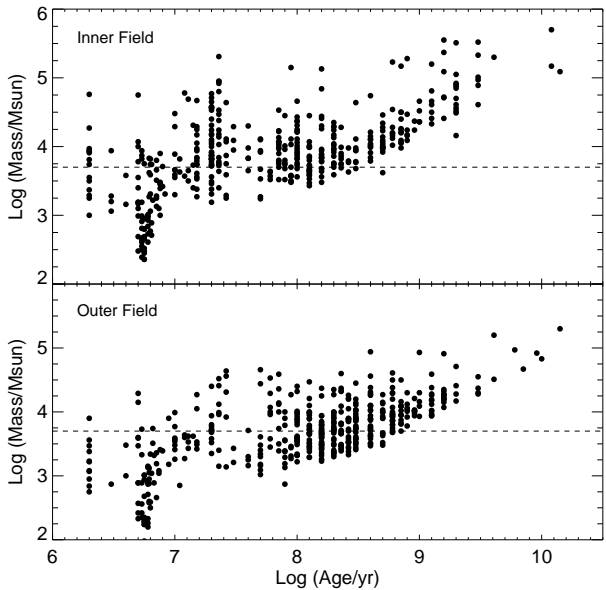}
\caption{The age vs. mass estimates for all clusters in the inner (top panel) and outer (bottom panel) fields.  The dashed lines represent the lower mass limit allowed in our sample in order to reduce the effects of stochastic IMF sampling on our dataset.}
\label{fig:age_mass}
\end{figure}

In the current work, we limit our analysis, when cluster parameters are fit, to those clusters that have masses in excess of $5\times10^3$\msun\ in order to minimise the effects of stochastic sampling of the stellar IMF that can severely affect the derived age and mass distributions (e.g. Ma\'{i}z Apell\'{a}niz 2009; Fouesneau \& Lancon, 2010; Silva-Villa \& Larsen 2011 ). Additionally, we do not include the inner 450~pc (in projection) around the galaxy nucleus as the star-formation rate may vary there (e.g. Harris et al.~2001) and the detection limit is significantly worse than in the outer regions.  After applying these mass and position limits our final sample contains 381 and 370 clusters in the inner and outer fields, respectively.  The age/mass distributions for these clusters are shown in Fig.~\ref{fig:age_mass}.

\subsection{Comparison with previous studies}
\label{sec:comparison}

In B11 we compared the derived ages for a sample of 48 clusters in common between the current work and that of C10 (and published in W11).  Overall, the agreement is excellent, with only a handful ($<8\%$) of clusters showing age differences larger than 0.5 dex.

Our sample selection differs notably from previous studies of extragalactic cluster populations, which impacts some of the conclusions reached in the present work relative to what has been reported earlier.  The cluster population in the inner field has been studied in detail in C10, and in this section we compare the samples, and discuss how the different samples influence the results.  There are two main differences between the sample in the current work and the one presented in C10.  The first, is the use of visual inspection of the images, and the inclusion of only centrally concentrated compact and symmetric sources.  The second difference, is that we do not include the inner portions of the galaxy (within $\sim450$~pc of the galactic centre in projection). The effect of these two differences are discussed in turn below.  We note that the comparison below is restricted to the inner field only, as it is the only field analysed by C10.

\subsubsection{The effect of sample selection}

%\fbox{mention C10 dual approach}

\begin{figure*}
\centering
\includegraphics[width=12cm]{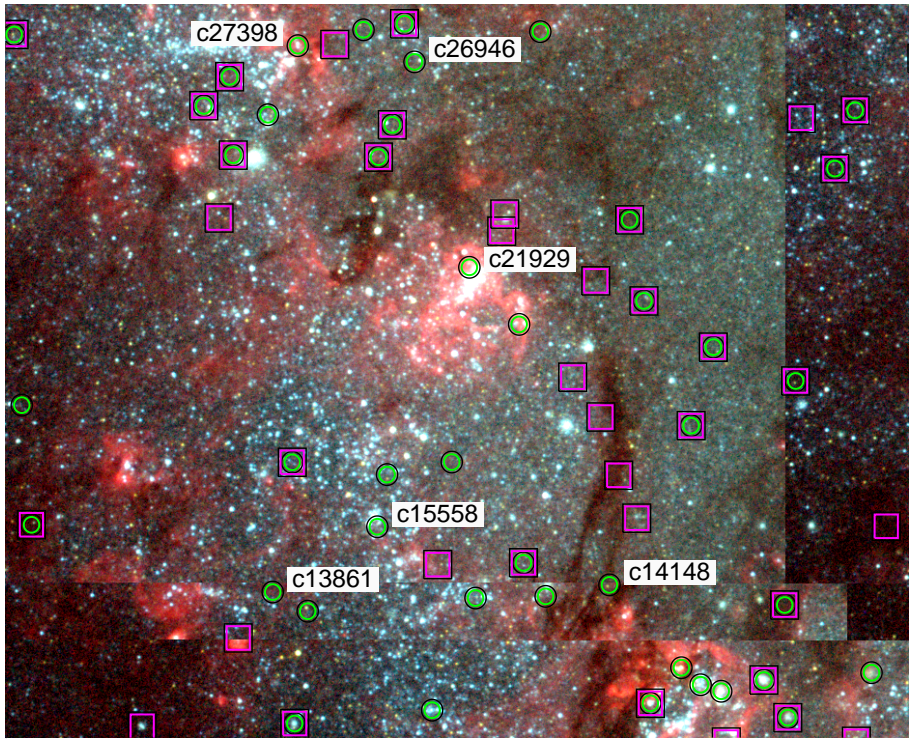}
\caption{A colour composite image of a $\sim815$~pc by $\sim665$~pc region in the inner field.  The B, V and H$\alpha$ images are shows as blue, green and red, respectively.  Sources from the C10 sample are shown as (green) circles while clusters in the present paper are shown as (magenta) squares.  Additionally, six sources from the C10 sample that are not in the current sample, that are shown in detail in Fig.~\ref{fig:example_chandar}, are labelled.}
\label{fig:region}
\end{figure*}

\begin{figure}
\centering
\includegraphics[width=2.7cm]{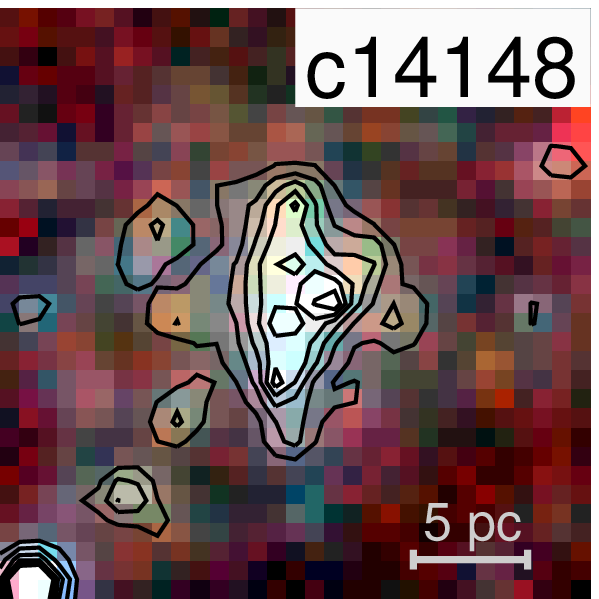}
\includegraphics[width=2.7cm]{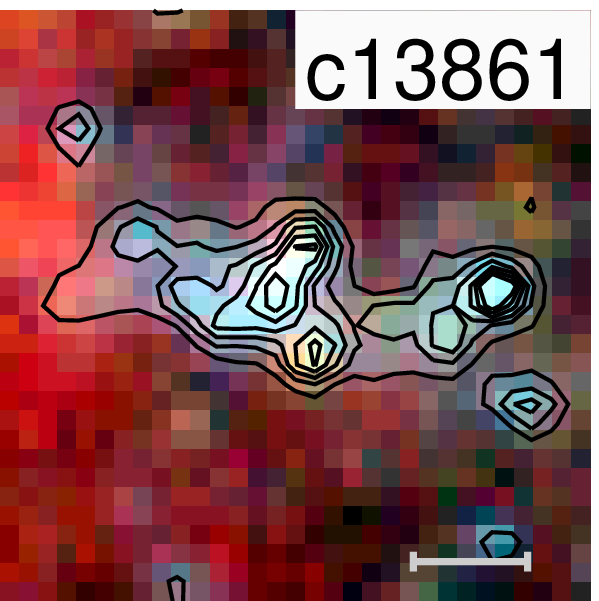}
\includegraphics[width=2.7cm]{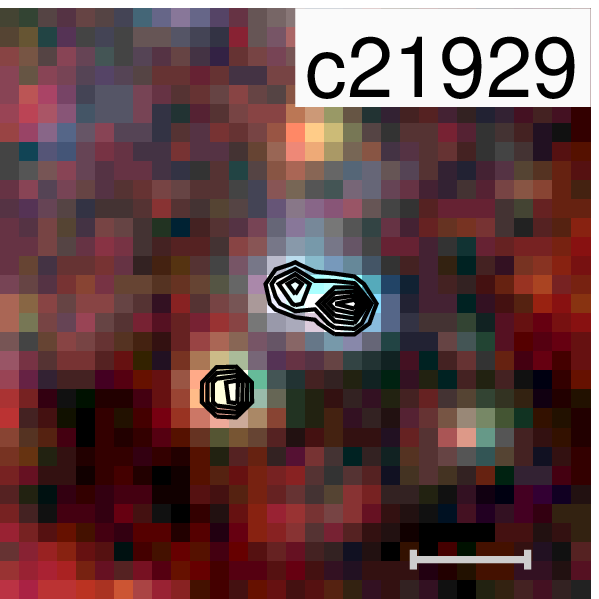}
\includegraphics[width=2.7cm]{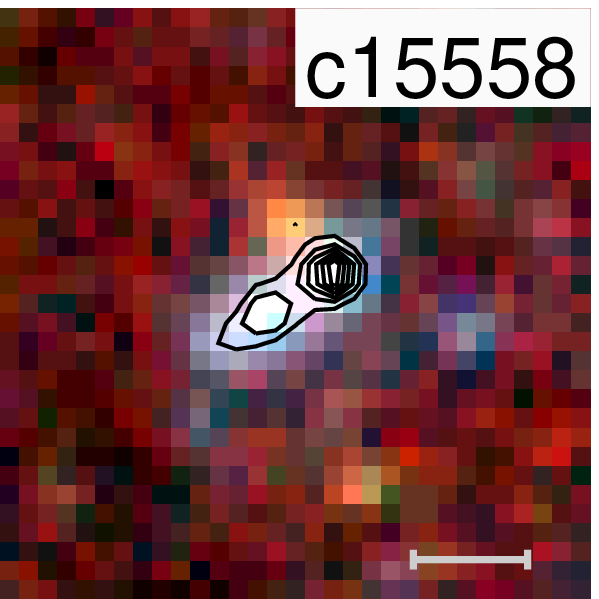}
\includegraphics[width=2.7cm]{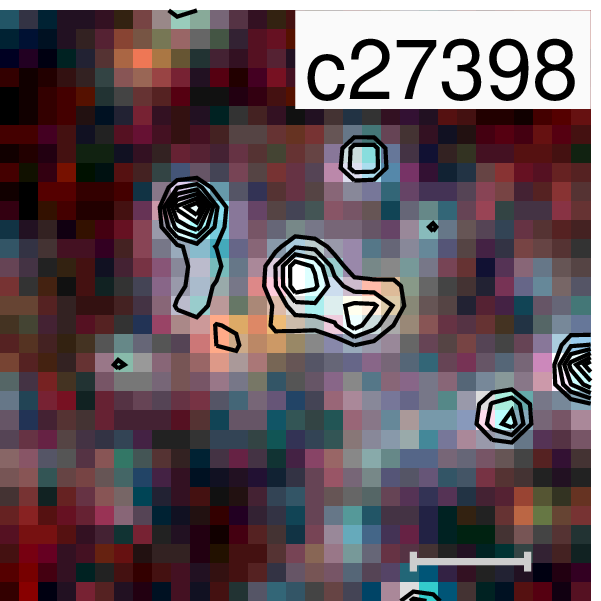}
\includegraphics[width=2.7cm]{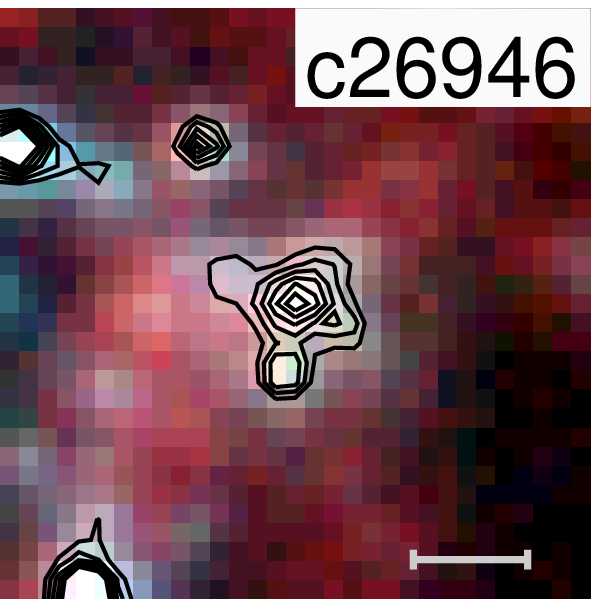}
\includegraphics[width=2.7cm]{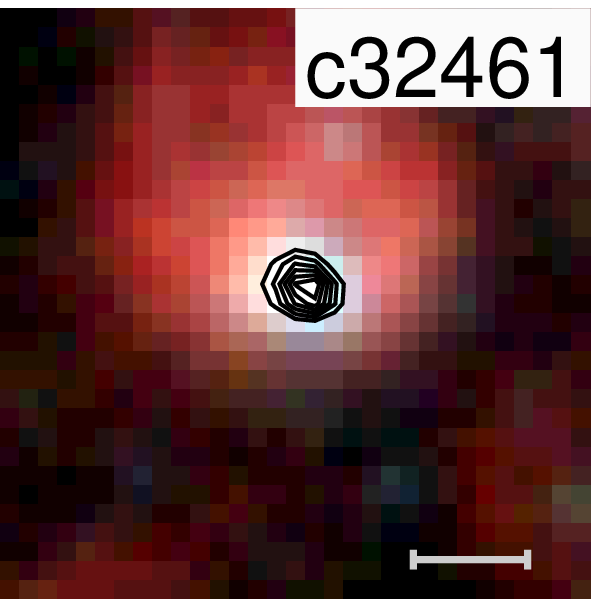}
\includegraphics[width=2.7cm]{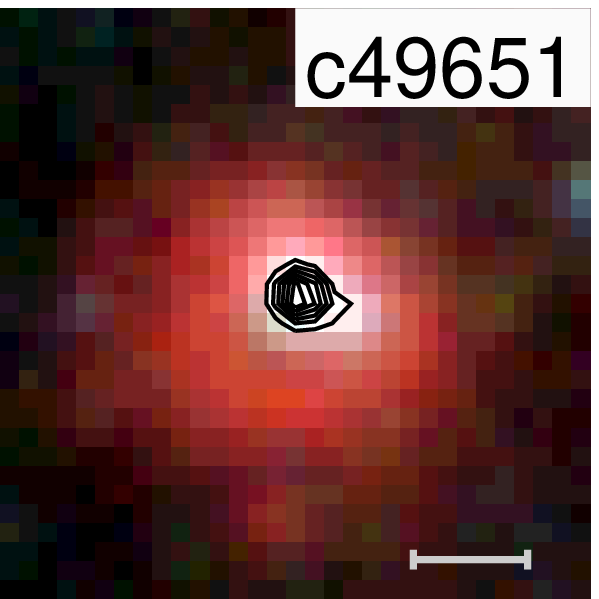}
\includegraphics[width=2.7cm]{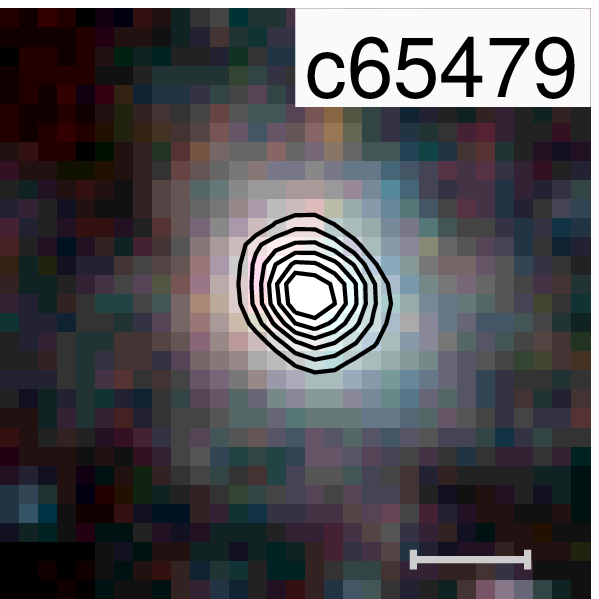}

\caption{{\bf Top and middle rows:} Examples of sources included in C10 that are not considered ``clusters" in this study, with each image centred on the source in the C10 catalogue.  The background shows a colour composite image (26~pc on a side) with B, V and H$\alpha$ represented by blue, green and red, respectively.  The contours denote V-band flux levels at 10\% steps from the maximum level of the source in question.  The bar in the lower-right corner of each panel represents 5~pc.  The ID from C10 is given in the upper right of each panel. {\bf Bottom row:} A sample of sources in the C10 catalogue that are also included in the present work.}
\label{fig:example_chandar}
\end{figure}

In Fig.~\ref{fig:region} we show a colour image of a section of the south-east corner of the inner field, covering a portion of the inner spiral arm.  Additionally, we show all of the source candidates from the current work (magenta squares) and the C10 catalogue (green circles).  While a number of sources appear in both catalogues, it is clear that many do not.  In order to highlight the differences, we show a blowup of candidate sources in the C10 catalogue that do not appear in the present work in Fig.~\ref{fig:example_chandar}, along with V-band flux contours (in steps of 10\% from the maximum level of the source).  Each of these sources fails to meet our selection criterion based on morphology, hence would likely fall into the associations category and not clusters based on the GPZ11 definition, if a characteristic size could be defined, which is not possible due to the complicated morphologies.  Physically, some part of each of these associations may become bound and will eventually form a cluster, however it is impossible to tell the fraction (if any) from imaging alone.  Some regions, e.g. c14148 or c13861, clearly do not fulfill our symmetry criterion.  Part of other regions, e.g. c27398 or c26946, may be defined as a cluster, depending on how exactly the selection is done.  This shows, in part, the subjectivity that enters the source selection process.

\begin{figure}
\centering
\includegraphics[width=8cm]{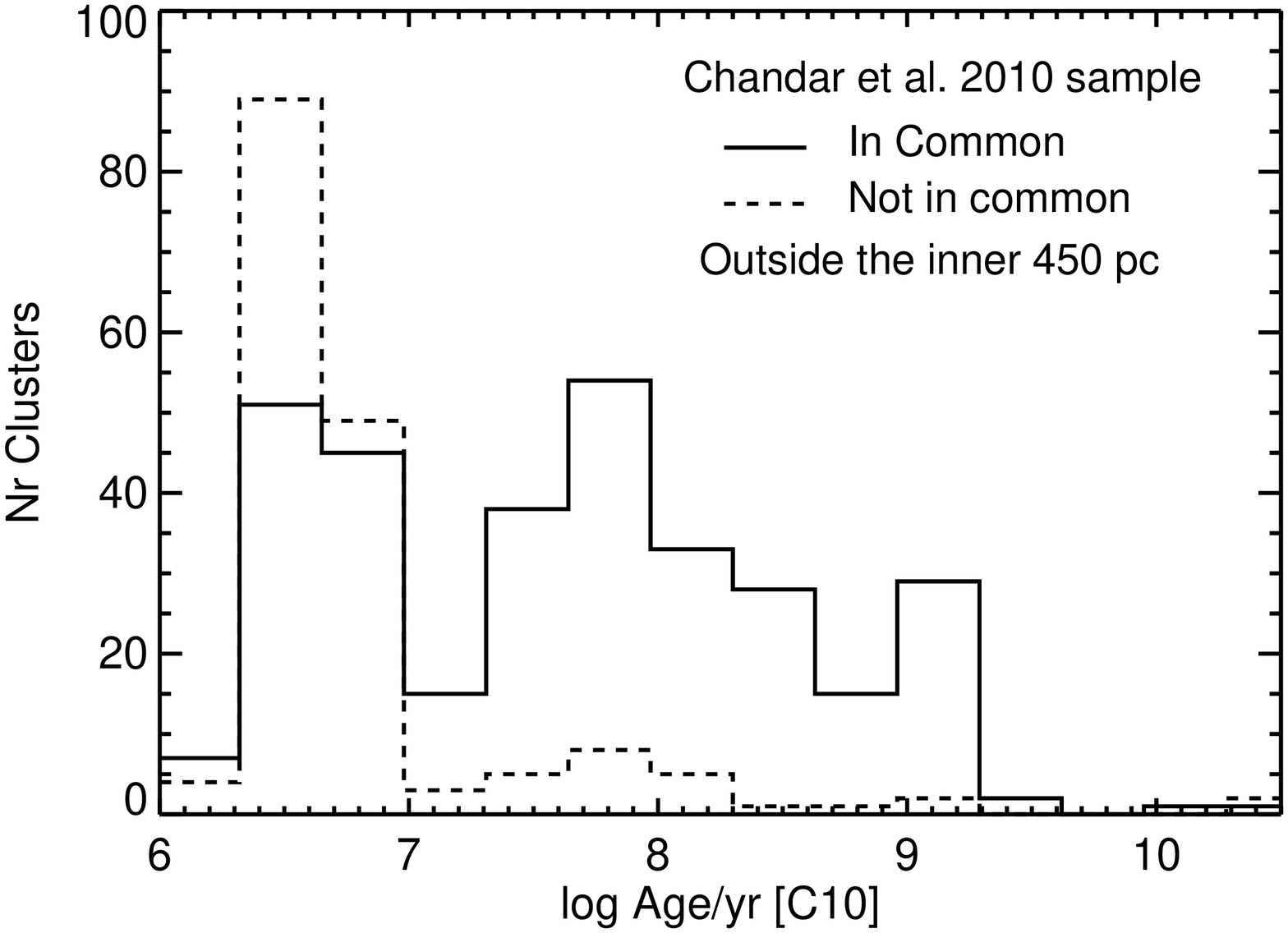}
\includegraphics[width=8cm]{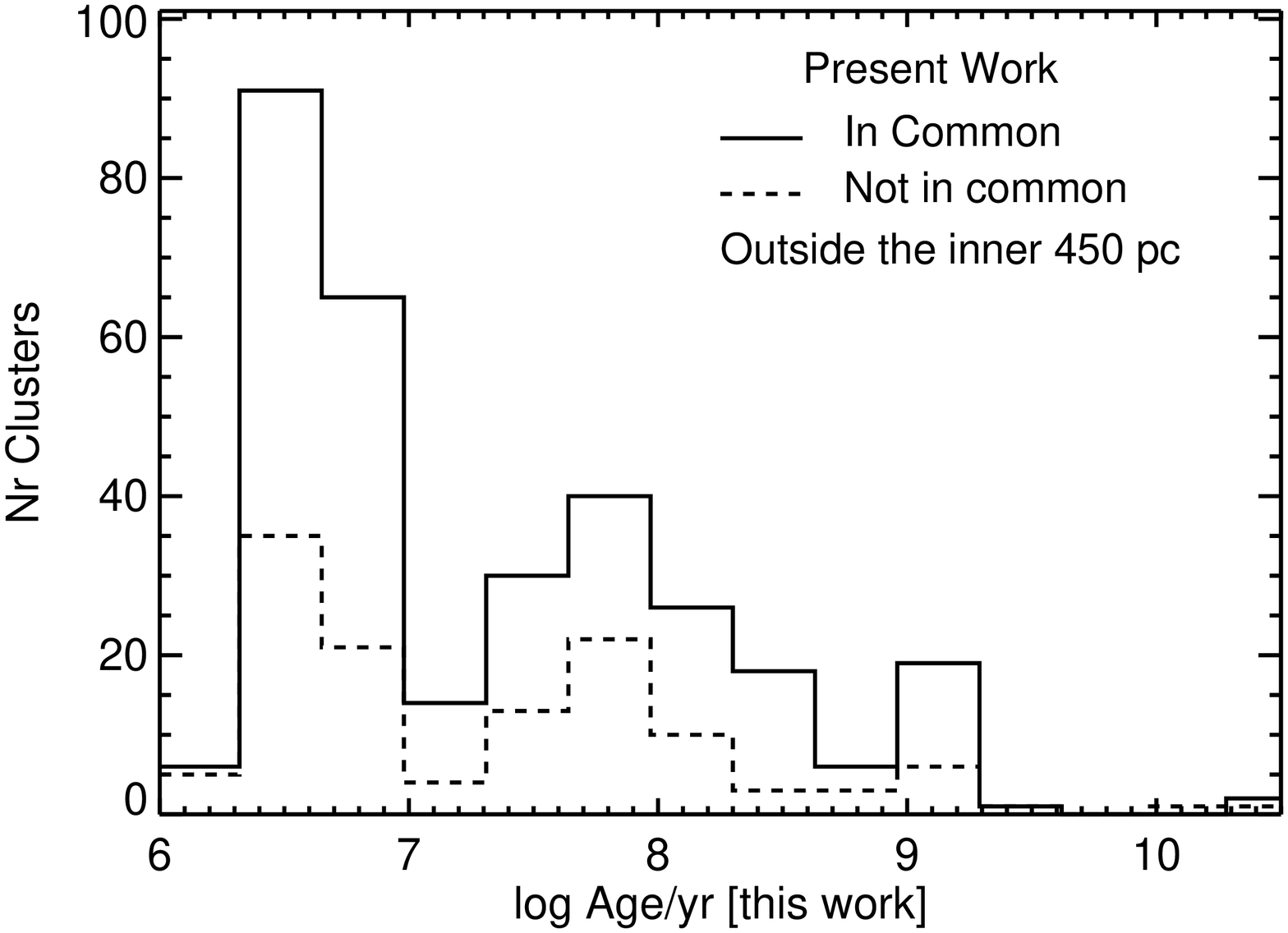}
\caption{Histogram of the ages of clusters in C10 and the current work, separated into those in common between the two studies (solid lines), and those that are not (dashed lines).  {\bf Top Panel:} From the catalogue of C10.  {\bf Bottom Panel:} From the catalogue of the current work.  All ages are taken from the respective works.  The main difference between the top and bottom panels is that a large fraction of the youngest ($<10$~Myr) clusters in the C10 catalogue are not in our catalogue.  This is likely due to the differing definitions of a ``clusters" between the two works (see Fig.~\ref{fig:example_chandar}).  This comparison includes all objects that appear in the respective works, independent of mass.  Only objects outside the central region (outside $\sim450$~pc) are included.}
\label{fig:age_dist_comp}
\end{figure}

However, we expect such subjective biases to be mainly applicable to the younger regions, since clusters will become more uniform (smooth) as they dynamically evolve (as the substructure is removed, e.g., Bastian et al.~2009b) and associations will dissolve into the background.  In Fig.~\ref{fig:age_dist_comp} we show the distribution of ages for each of the catalogues, divided into sources that appear in both catalogues (solid lines) and those that appear only in one of the two (dashed lines).  Only sources outside of the inner $450$~pc are included from each catalogue.  In the top panel, we show the age distribution from the C10 catalogue, while the bottom panel shows the age distribution in the present work.  All sources in the respective catalogues have been used in the figure, i.e. no cut on mass has been applied (contrary to the rest of the current paper).

Focussing on the top panel, we see that the distribution of ages of the sources in common between the catalogues does not have a strong dependence on age (i.e. being dominated by only the youngest or oldest clusters).  However, the sources in the C10 catalogue that do not appear in the present catalogue are strongly peaked at young ages ($<10$~Myr), confirming that the differing selection criteria between the two studies is largely limited to the youngest ages.  This confirms the expectations discussed above.

In the bottom panel, we see that the age distribution of the sample that is in common with C10 and the sample not in common do not show such a difference.  From these two panels we conclude that the two catalogues differ mainly in the youngest ages.  

Due to this difference, we will largely limit our analysis to clusters with ages in excess of 10~Myr.  However, we will return to this difference when discussing the fraction of star-formation that happens in clusters as well as the role of infant mortality in \S~\ref{sec:discussion}, both of which are heavily affected by the sample selection.

We emphasise the subjective nature of the differences between the two studies.  The present sample clearly misses some fraction of clusters that form from the densest parts of associations, while the C10 sample includes a larger number of young objects for which we cannot be sure whether they are (or will remain) bound.  This shows the limitation of extragalactic studies of young cluster populations.  Finally, we note that C10 used a ``dual approach", selecting sources with automated methods (the catalogue we compare with here) as well as by eye.  The authors find good agreement between the two selection techniques at the 80\% level, although the ``by eye" selection criteria were not specified, and most/all of the groups shown in Fig.~\ref{fig:example_chandar} were included in their ``by eye" catalogue.

\subsubsection{The effect of not including the inner galactic regions}

Another difference between our sample selection and that of C10 is that we exclude the innermost regions of the galaxy ($<450$~pc).  We exclude this for two reasons.  The first, is that the ongoing starburst in the centre of the galaxy means that the star-formation rate within this region may not have been constant over the past few hundred Myr (e.g. Harris et al.~2001; Fathi et al.~2008).  Secondly, due to the increased surface brightness, complicated dust/gas structure, and heavy extinction, the detection limit within this region is significantly worse than in other portions of the images.   Since clusters fade strongly after the first few Myr of their lives, continuing as a near power-law decrease in brightness for the next few hundred Myr, this will strongly bias the sample to only detecting the young brightest clusters.

\begin{figure}
\centering
\includegraphics[width=8cm]{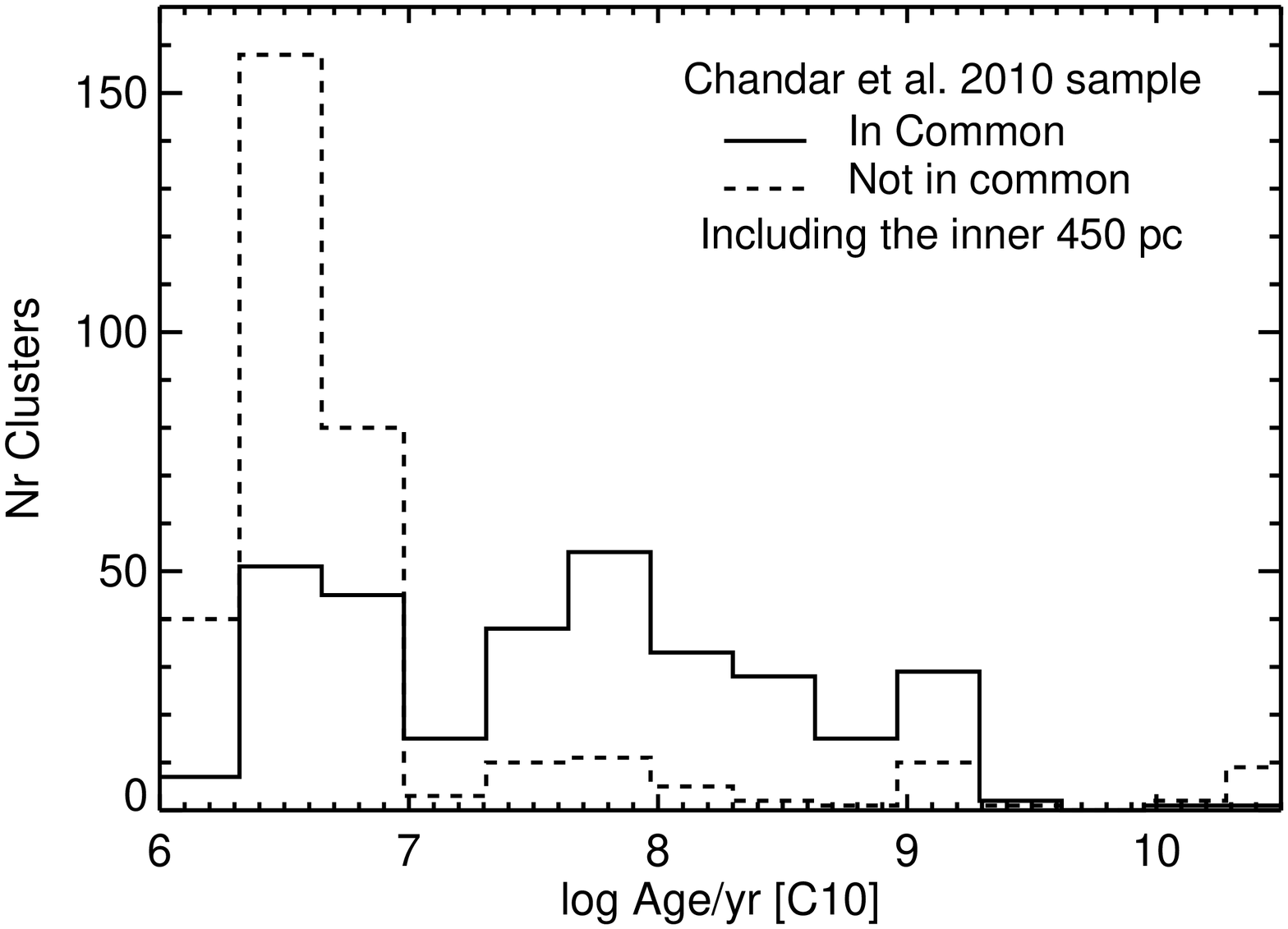}
\caption{The same as Fig.~\ref{fig:age_dist_comp}, except now all objects in the C10 are included (including those within the inner $\sim450$~pc.).  As in Fig.~\ref{fig:age_dist_comp}, we see that the two catalogues differ most at the youngest ages.}
\label{fig:age_dist_comp2}
\end{figure}

In Fig.~\ref{fig:age_dist_comp2}, we show the age distribution of the full C10 sample (i.e. including the inner regions) split into sources found in the current catalogue (solid line) and absent in our catalogue (dashed line).  As was seen in the proceeding section, but now amplified even more, the C10 sample contains a large number of young sources that are not included in our sample.

The removal of the inner 450~pc of the galaxy from our sample should not affect our results if the cluster age and mass distributions are universal (i.e., independent of location, e.g., Whitmore et al. 2007).  We will however still have a large range of environments to sample to search for the environmental dependence of cluster disruption as well as any truncation that may exist in the mass function (above our detection limit).

In \S~\ref{sec:disruption} we will show that the age distributions, beyond the first 10~Myr, of the C10 and present catalogues agree quite well in the inner field, giving confidence in the ability to study extragalactic cluster populations.  We will then be in a position to compare the inner and outer fields to gain insight into the influence of the environment in the cluster formation and destruction processes.

%\begin{figure}
%\includegraphics[width=8.5cm]{Figs/age_comparison.ps}
%\caption{A comparison of the derived ages using the {\it 3DEF} method vs. those derived by C10 and reported in W11.  Solid points represent clusters where the two methods agree within 0.5 dex while red points indicate clusters where the derived ages disagree by more than 0.5 dex.  Blue boxes denote clusters where H$\alpha$ morphology and surface brightness variations indicate ages in excess of 10~Myr (W11).}
%\label{fig:age_comp}
%\end{figure} 

\subsection{$\Pi$ distribution}
\label{sec:pi}

As discussed above, it is not trivial to distinguish bound stellar groups (i.e. clusters) from unbound aggregates (i.e. associations) for young stellar systems.  At older ages, i.e. a few tens of Myrs, this distinction is much easier to make due to the dispersal and disappearance of unbound associations into the field.  However, at young ages, stellar groups often appear to be hierarchical or filamentary (e.g. Gutermuth et al.~2005; 2009, Allen et al.~2007) making it difficult to distinguish groups that are gravitationally bound vs. those that will disperse.  Even in fully resolved stellar systems it is non-trivial to define which stars belong to clusters (e.g. Bressert et al.~2010). 

%In order to address this uncertainty, GPZ11 have suggested an objective way to distinguish between clusters and associations.  The method is based on a dynamical definition, in that bound clusters will have (stellar) ages larger than a dynamical age (e.g. a crossing time).  The authors introduce a measurement, $\Pi$, defined as $t_{\rm age}/t_{\rm cross}$, where $t_{\rm age}$ is the stellar age of the system and $t_{\rm cross}$ is the crossing time, which can be used to classify stellar groups. If $\Pi$ is greater than 1 then the group is a cluster, whereas if $\Pi < 1$ then the group is an association.   For very young objects (i.e. younger than a crossing time) it is impossible to determine if a group is bound, hence it is not possible to define clusters until they are dynamically evolved.

In order to address this uncertainty, we have applied the GPZ11 definition of a cluster, namely that the age of the system is greater than the instantaneous crossing time, parameterised by $\Pi$.  Without kinematic information on each of the stars within a cluster it is impossible to measure $t_{\rm cross}$.  However, if one assumes that the clusters are in virial equilibrium, then the velocity dispersion can be estimated by combining the estimated mass and radius of each group.  This in turn can be compared to the derived age of the cluster (through SSP modelling) to estimate whether the cluster is bound (see GPZ11). 

\begin{figure}
\includegraphics[width=8.5cm]{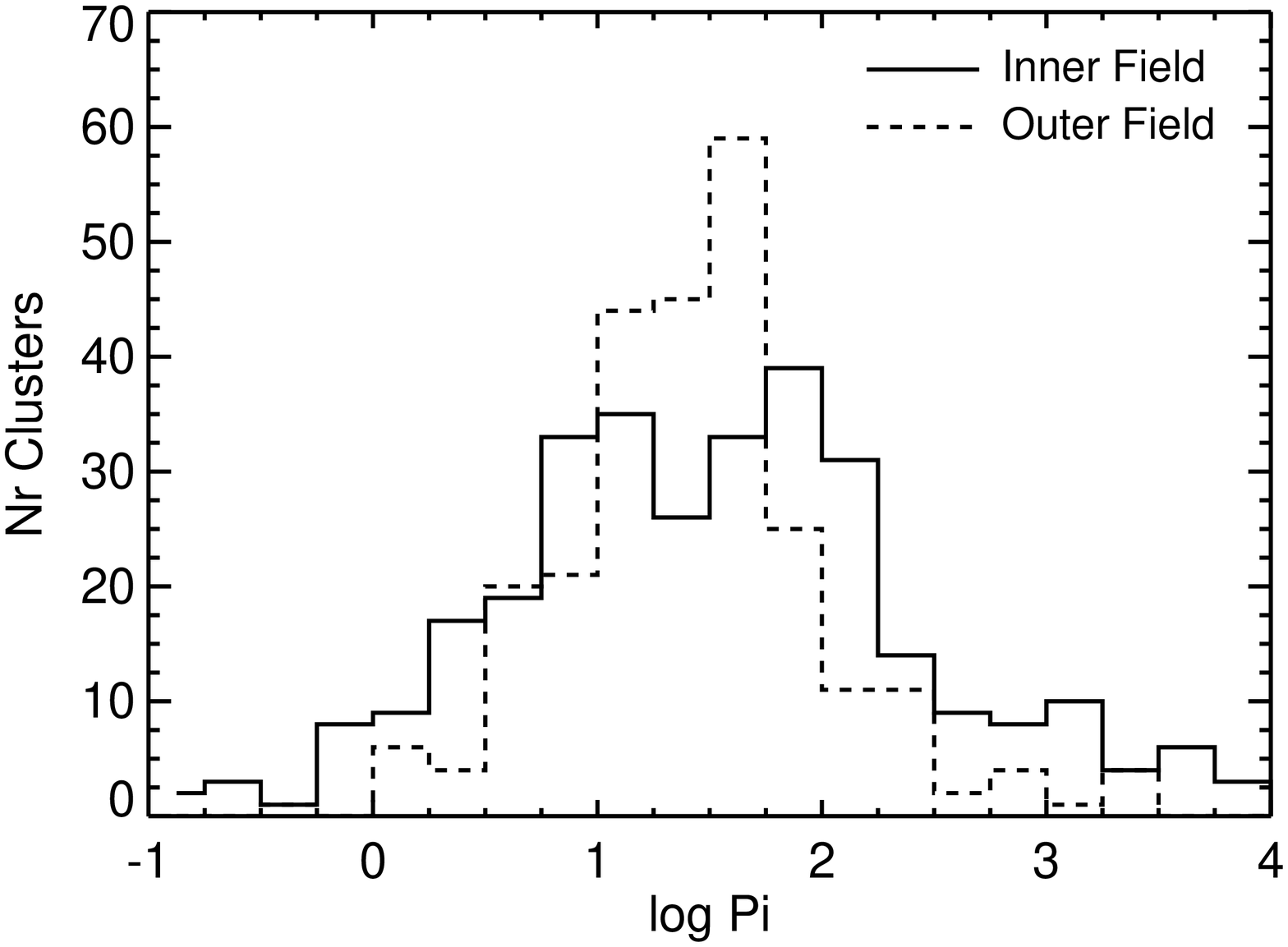}
\caption{The distribution of the derived $\Pi$ ($t_{\rm age}/t_{\rm cross}$) values of the observed clusters, split into each of the respective fields.  A value of log$(\Pi)=0$ separates clusters (log$(\Pi) > 0$) from associations (log$(\Pi) < 1$).
}
\label{fig:pi_dist}
\end{figure}

Figure~\ref{fig:pi_dist} shows the distribution of $\Pi$ for the clusters in the inner (solid line) and outer (dashed line) fields.  Note that the vast majority of candidates fulfil the definition of a cluster as defined by GPZ11.  We include all objects that appear in Fig.~\ref{fig:pi_dist} in the subsequent analysis, even those with log($\Pi$) $< 0$.  However, we note that our selection results in a sample that includes
  very few objects with log ($\Pi$) $< 0$.

\subsection{Completeness limits and the luminosity function}
\label{sec:lf}

In order to use the $\Pi$ definition of a cluster, accurate sizes are needed.  This, in turn, requires high enough signal to noise in order to obtain reliable profile fits to the individual clusters.  In order to determine the limiting magnitude for reliable fits, we employed a similar technique to that used by Silva-Villa \& Larsen~(2011).  Artificial clusters were created by stochastically sampling the stellar initial mass function (Kroupa~2001), adopting a total stellar mass of $1\times10^4$\msun, and an age of 50~Myr.  Luminosities and colours were assigned to each star by applying the Marigo et al.~(2008) isochrones.  Images of the artificial cluster were then made by convolving them with the empirically derived PSF and the WFC3 {\it UVIS} properties.  Clusters were created with effective radii of 3 and 5 parsecs, following a King~(1962) profile.  One hundred realisations were made for each effective radius and added to the images.  We focussed our attention on the V-band images since the selection was based largely on those.

The clusters were then scaled to various luminosities to test for completeness.  We found that clusters in the inter-arm regions of M83, where the background is low, were readily detected fainter than V = 23~mag.  However, in the crowded and bright regions in spiral arms, the detection limit was notably brighter.  We adopted detection limits of V=22~mag and V=22.5~mag in the inner and outer fields, respectively.  These conservative limits were found by requiring that the artificial clusters could be identified and the properties robustly determined even in crowded, high background regions.  We note that some fainter clusters appear in our catalogue due to different aperture sizes during the selection stage and the final photometry.

\begin{figure}
\includegraphics[width=8.5cm]{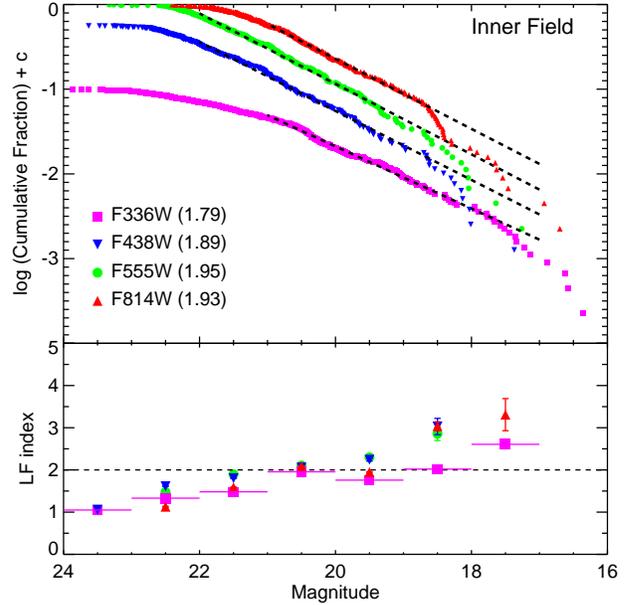}
\caption{{\bf Top:} The cumulative luminosity function of clusters in M83 in the four main filters used in this work. The numbers in parenthesis correspond to the measured index (assuming $NdL \propto L^{-\alpha}dL$) over the range indicated by the dotted lines over the data.  {\bf Bottom:} The derived index as a function of brightness in each magnitude interval (only for intervals with five or more clusters).  Note the steepening in each of the bands at bright magnitudes.
}
\label{fig:lf1}
\end{figure}

\begin{figure}
\includegraphics[width=8.5cm]{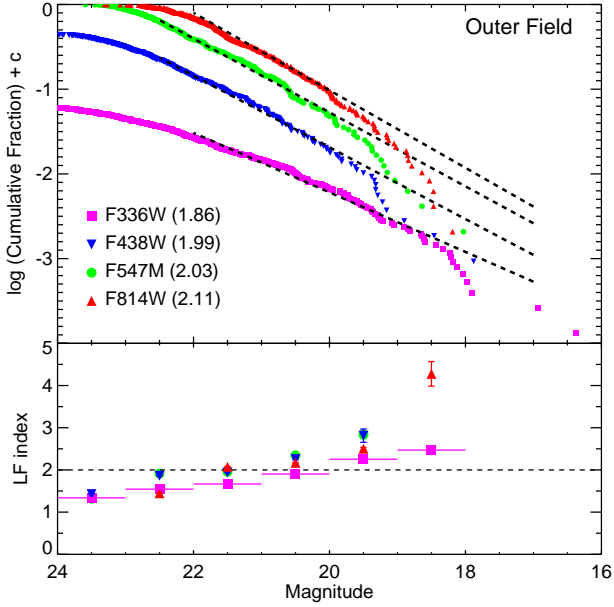}
\caption{The same as Fig.~\ref{fig:lf1} but now for the outer field.
}
\label{fig:lf2}
\end{figure}

The resulting luminosity function (LF) of clusters for the inner and outer fields are shown in Figs.~\ref{fig:lf1} \& \ref{fig:lf2}, respectively.  The LF are shown as cumulative functions and have been offset for clarity.  In each figure, the I, V, B, and U-bands are shown from top to bottom in the upper panels.  The dashed lines in the top panels show the maximum likelihood fit over the range covered by the line, with measured value shown in parenthesis, assuming the form $dN/dL \propto L^{-\alpha}$.  The lower panels show linear fits (made in the cumulative distribution) to the distribution in 1.0 magnitude bins.  As in previous studies, we find that $\alpha = 2.0$ provides a good fit to the data over most of the observable range (de Grijs et al.~2003).  We note that associations with sizes of several tens or even hundreds of parsecs also show the same luminosity function (Bastian et al.~2007), so a power-law with an index of $-2$ appears to be a general result of star-formation and not specific to clusters.  This index likely is due to the scale-free and hierarchical nature of the ISM from which stars form (Elmegreen~2002).

However, we see a steepening of the LF above a V-band magnitude of 19 and 19.5 (corresponding to an absolute V-band magnitude of -9.3 and -8.8) in the inner and outer fields, respectively.  This steepening has been observed in the LF at these magnitudes of cluster populations in a sample of galaxies (Larsen~2002; Gieles et al.~2006a,b; PZMG10) and has been interpreted as being due to a truncation in the underlying mass function at high masses.    The underlying mass function of clusters and the possible truncation at high masses will be further discussed in \S~\ref{sec:mass_func}.

\subsection{Star-formation history of M83}
\label{sec:sfh}

When using observations of cluster populations in order to understand cluster disruption, it is important to take into account changes in the star-formation history (SFH) of the galaxy during the period under study.  For example, if the SFH of a galaxy has been increasing, this will have the same observational signature as mass independent disruption (i.e. more young clusters will be seen in both cases).  Hence,  cluster disruption will be over-predicted in such a case if the SFH is not taken into account (e.g. Bastian et al.~2009a).  This type of degeneracy has also been discussed in Pellerin et al.~(2010) and Annibali et al.~(2011).

Since we are studying the cluster population within a single galaxy, separated into two radial bins, it is important to check if the SFH between the two fields has been the same.  In particular, we need to check if there is a radial dependence in the SFH (outside the inner $\sim450$~pc).  In order to check this, we have used the resolved stellar population analysis carried out by Silva-Villa \& Larsen~(2011) in M83, based on two pointings of the Advanced Camera for Surveys (ACS) onboard HST.  Their two fields are centred at nearly the same galactocentric radii as the WFC3 observations used in the present work (centred at $\sim2.9$ and $4.8$~kpc  from the galactic centre in the ACS study and 2.5~kpc and 4.75~kpc in the present study).  Silva-Villa \& Larsen~(2011) have derived the SFH within these fields for the past $10-100$~Myr.  

\begin{figure}
\includegraphics[width=8.5cm]{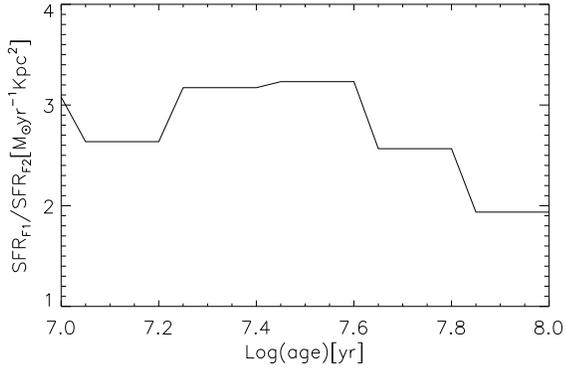}
\caption{The ratio of the SFHs derived from the two ACS pointings discussed in Silva-Villa \& Larsen~(2011).  While the SFH has been higher in the inner field (as expected) the ratio of the SFH between the two fields has been largely constant, meaning that any differences between the cluster populations in the two fields is likely due to disruption and not differences in their formation histories (unless if the SFH of M83 varies azimuthally).
}
\label{fig:sfh}
\end{figure}

The ratio of the derived SFHs of the two fields is shown in Fig.~\ref{fig:sfh}.  The comparison of the SFHs in the two ACS fields centred at about the same galactocentric distances as our WFC3 fields shows that the ratio of the SFRs in the two ACS fields has remained relatively constant (within $\sim50$\%) over the past 100 Myr. However, it is important to note that the outer ACS field is located quite far from our outer WFC3 field, and we do not have strong constraints on any azimuthal variations in the SFHs. We therefore do not have very strong constraints on the relative SFHs of the two WFC3 fields.

%While the SFH has been higher in the inner field (as expected) the ratio of the SFH between the two fields has been largely constant, meaning that any differences between the cluster populations in the two fields is likely due to disruption and not to differences in their formation histories.  There does appear to be a small increase in the SFH ratio between the two fields ($<40$\%), which is much smaller than the factor of 2.5 needed to explain the observed differences between the cluster populations (discussed in \S~\ref{sec:disruption}).

\section{Empirical age, mass and size distributions}
\label{sec:results}

Using the derived properties of the 751 clusters in our sample, we can now investigate properties of the cluster population as a whole.  In particular, we are interested in properties that vary as a function of environment, which will give the best constraints on cluster formation, evolution and destruction.

Here, we focus on 1) the cluster radius (size) distribution in comparison to the results of other cluster populations available in the literature, 2) the form of the mass function, in particular on any possible truncation at high masses and 3) the age distribution, which will be used to study cluster disruption.  Due to interdependencies between the age and mass distributions as well as completeness issues, issues 2) and 3) will be modelled simultaneously.

\subsection{Age distributions}
\label{sec:age-dist}

As noted in B11, the distribution of cluster ages appears to be different between the two fields.  This was based on the estimated ages/masses, but could also visually be seen by looking at the colour distributions of the populations in the two fields.  Both populations had similar mean/median V-I colours, but the outer field had a significantly redder U-B colour.  When comparing the distribution to SSP models, it was shown that extinction could not be the cause, as the inner field is expected to have more extinction, and then the V-I colours between the two fields should also be different.

\begin{figure}
\includegraphics[width=8.5cm]{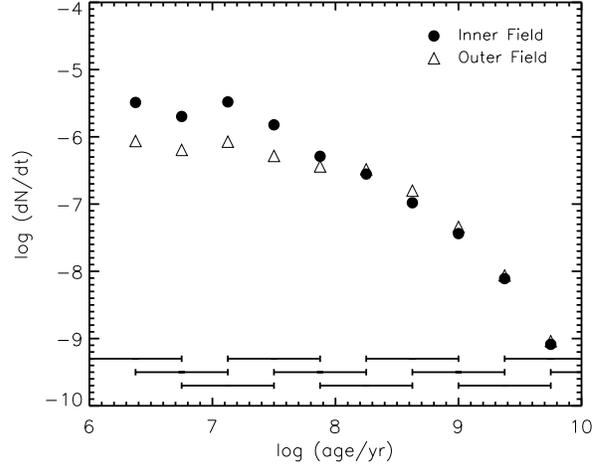}
\caption{The age distribution (number of clusters per unit time) for the inner (filled circles) and outer (open triangles) fields.  Only clusters with masses above $5\times10^3$~\msun\ are used.  The bars in the bottom of the panel show the age bins used for the plot.  In order to minimise the effect of spurious features in the age distribution, a series of overlapping bins are used.  Ages above $\sim1$~Gyr are highly uncertain and the sample is affected by completeness issues above this age.
}
\label{fig:age-dist}
\end{figure}

In Fig.~\ref{fig:age-dist} we show the age distribution (number of clusters per unit time) for the inner and outer fields.  The sample used is mass limited, with a lower limit of $5\times10^3$~\msun.  At older ages both samples are affected by the completeness limit (i.e. the drop in dN/dt at older ages).  We find that for most mass cuts, the outer field has a median age that is $2 - 3$ times higher than the inner field.

In order to avoid incompleteness effects, we will use the mass distributions in \S~\ref{sec:disruption} in order to study the effect of disruption on the two fields.  However, the age distribution of clusters in the outer field is largely flat up to an age of $\sim100$~Myr, indicating that cluster disruption has not significantly affected the distribution (e.g. Boutloukos \& Lamers~2003).

\subsection{Size distribution}
\label{sec:size}

\begin{figure}
\includegraphics[width=8.5cm]{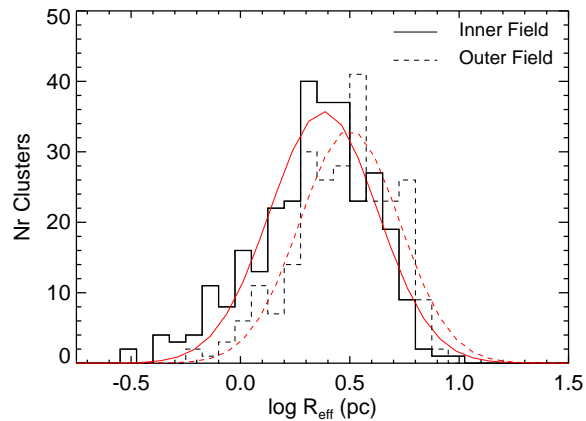}
\caption{The distribution of the derived effective radii (\reff) of the clusters, split into the respective fields.
The thin (red) lines represent the best fitting Gaussian distributions.  The clusters in the outer field have a peak at slightly larger radii ($\sim2.6$~pc) compared to the inner field ($\sim2.3$~pc).  This difference is likely due to an underlying age/size relation (see Fig.~\ref{fig:age_reff}).}
\label{fig:size_dist}
\end{figure}

Next, we turn to the effective radius (or size) distribution of the clusters in our sample.  In Fig.~\ref{fig:size_dist} we show the distribution of measured radii in the V-band for each of the two fields.  For the fits, we have adopted an EFF profile (Elson, Fall \& Freeman~1987) with index, $\gamma/2=1.5$, and a fitting radius of five pixels.  Additionally, we show the best fitting log-normal functions to each of the datasets, which are shown as light (red) lines.  Both distributions appear to be reasonably fit with a lognormal function, although we note that we are not sensitive to very small clusters ($<0.5$~pc) as they will appear as unresolved on the HST images. The size distributions peak at $\sim2.3$~pc and $\sim2.6$~pc for the inner and outer field, respectively.  The results are unchanged if the B or I-bands were used instead.

These results are very similar to that found by Scheepmaker et al.~(2007) for the cluster population of M51 (with a median effective radius of 2.1~pc and a similar lognormal form) and for clusters in M31 (Vansevi{\v c}ius et al.~2009 - 2.1~pc when binned in the same way as done here).  It is also close to the median value of 3.2~pc for ``blue clusters" in M101 derived by Barmby et al.~(2006).  The median size, however, is much smaller than that found by Mayya et al.~(2008) for massive ($>10^{5}$\msun) clusters in M82 (with a median of $\sim10$~pc).  However, studies of individual massive clusters in M82 result in much smaller sizes (with a median of $\sim1.8$~pc; e.g.,~ McCrady, Gilbert, \& Graham~2003;  Smith et al.~2006).

\begin{figure}
\includegraphics[width=8.5cm]{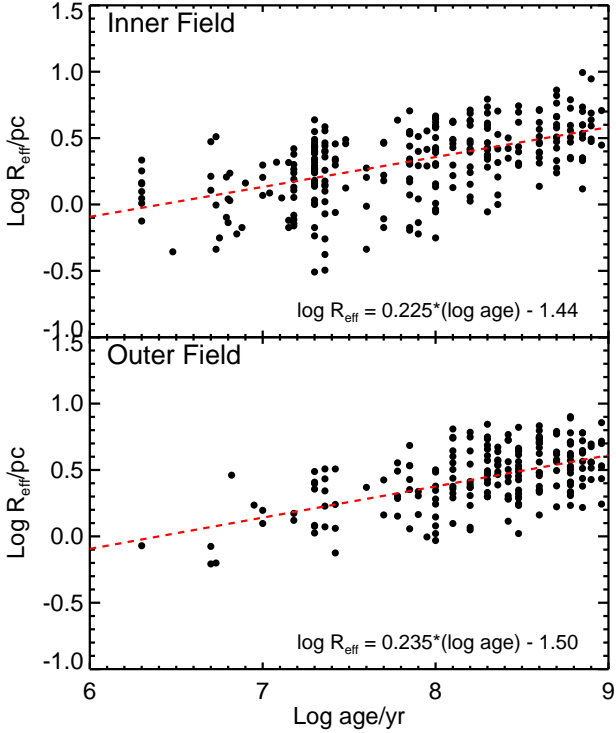}
\caption{The age vs. size relation of clusters in the inner (top panel) and outer (bottom panel) fields.  The best linear fit to the data is shown as a dashed (red) line.  Note the similar expansion in both fields, implying that cluster expansion is driven by internal mechanisms or by a similar observational bias.}
\label{fig:age_reff}
\end{figure}

Scheepmaker et al.~(2007) found an interesting trend between cluster radius and galactocentric distance in that clusters further from the galactic centre were larger on average.  A similar trend is seen in Fig.~\ref{fig:size_dist} for the clusters in M83.  In order to see whether this reflects different formation mechanisms, we have looked for correlations of R$_{\rm eff}$ with other derived properties.  In Fig.~\ref{fig:age_reff} we show the observed relation between the size and age of clusters in the two fields.  Only clusters with masses above $5\times10^3$\msun\ were used.  Additionally, we show the best fitting linear relation for both fields (fit separately) as dashed (red) lines.  The resulting relations are also given in the panels.  Both fields show the same trend, which will be discussed in more detail below.  

As noted above, the outer field has a significantly older median age than the inner field (approximately a factor of 2 or 3 higher, depending on the mass cut).  Hence, simply due to the observed expansion (discussed below), we would expect clusters in the outer field to be larger, on average, than those in the inner field.  Quantitatively, based on the median age difference, we would expect the clusters to be $\sim0.4$~pc larger in the outer field.  This is quite similar to the observed difference of  $\sim0.3$~pc, suggesting that the observed size difference between the two fields is driven mainly (or completely) by the expansion of clusters and the difference in the average age between the two fields.

Another possibility of the driver behind the age/size relation would be an underlying age/mass relation.  Since we are not sensitive to old, low mass clusters, this could affect our size distributions.  In order to test this, we performed a bivariate fit on the data (again, limiting the analysis to clusters more massive than $5\times10^3$\msun\ and ages less than 1~Gyr).  We fit a function of the form: 
\begin{equation}
%$$  {\rm log (R_{eff})}  = a_1 \times {\rm log (age)} + a_2 \times {\rm log (mass)} + {\rm constant}$$.  
$$  {\rm R_{eff}}  \propto t^{a_1} M^{a_2}$$,% \times {\rm log (age)} + a_2 \times {\rm log (mass)} + {\rm constant}$$.  
\end{equation}
where t and M are the age and mass of a cluster, respectively.  The fit was carried out independently for each field.  For Field~1 we find: $a_1 = 0.22 \pm 0.04$ and $a_2 = 0.04 \pm 0.08$.  This shows that the observed increase in size with age, is not strongly affected by any mass relation.  For Field 2 we find $a_1 = 0.24 \pm 0.06$ and $a_2 = 0.22 \pm 0.15$.  Here the result is more ambiguous, again we do see an intrinsic relation between age and size, but also a weak relation between mass and size, albeit with large errors.

Given these results, we conclude that there is evidence for an increasing size as a function of age, qualitatively similar to that observed in other cluster populations (Mackey et al.~2003; Bastian et al.~2008).  This could be a physical effect, due to internal heating by stellar mass black holes, hard binaries and stellar evolution (e.g. Merrit et al.~2004; Mackey et al. 2007, 2008; Gieles et al.~2010), or gas expulsion (Goodwin \& Bastian~2006), or a combination of the above effects.  Alternatively, it could be a selection effect due to the assumption of a single profile to fit all clusters, i.e. ignoring mass segregation or the evolution of the profile shape (e.g. Gaburov \& Gieles~2008), or a stochastic effect with the profile of the younger clusters being dominated by a few very bright stars, which tends to result in smaller derived radii than the true value (Silva-Villa \& Larsen~2011).  A detailed study of the profile shapes and sizes of a sample of clusters in our catalogue will be presented in a forthcoming paper (Konstantopoulos et al.~in prep).

%\subsection{Cluster Expansion}
%\label{sec:expansion}

%Reff vs. Rcore

%much slower than seen in the rcore vs. age plot.

%Bastian et al. 2008
%Mackey et al. 2003

% It appears that the higher pressure/density central portions of galaxies preferentially contain more compact clusters.  Whether this is due to clusters forming in a more compact state in the inner regions of galaxies, or if the environment there is more disruptive to larger clusters cannot be answered with the present dataset.  If the answer is the former, this could have a significant effect on the derived metal abundance gradients in galaxies as derived by strong line emission measures (e.g. Ercolano, Bastian \& Stasi{\'n}ska~2007) or help constrain stellar cluster formation models (e.g.~Elmegreen~2008).  Extending this type of analysis to more extreme regions may shed light on this question.  As discussed below, this is just one property of a cluster population that appears to depend on distance from the galactic centre.

\subsection{The Cluster Mass Distribution}
\label{sec:mass_func}

One of the basic properties of a cluster population is the distribution of cluster masses.  Previous studies of extragalactic cluster populations have generally found a power-law type distribution with few high mass and many low mass clusters, namely ${\rm d}N/{\rm d}M \propto M^{-\beta}$.  Many studies have derived values of $\beta \sim 2$ (e.g. Zhang \& Fall~1999, de Grijs et al.~2003, Bik et al.~2003; McCrady \& Graham~2007) for various mass ranges, from a few thousand solar masses to $>10^6$\msun.  However, based on the luminosity distributions of cluster populations in a sample of spiral galaxies, Gieles et al.~(2006a,b) found evidence for a truncation in the distribution at the high mass end, which they approximated as a Schechter function (i.e. a power-law at the low mass end with an exponential cutoff at high masses).  Further evidence for a Schechter type distribution was presented by Bastian~(2008) based on the observed relation between the star formation rate of a galaxy and the luminosity of its brightest cluster.  Larsen~(2009) studied the relation between the age of a cluster and its luminosity in a sample of 16 galaxies and concluded that a Schechter function with a cutoff of a few times $10^5$\msun\ provided a good fit to the data.  Maschberger \& Kroupa~(2009) showed that the cluster population in the LMC has a truncation in the mass function at $\sim7\times10^{4}$\msun.  Finally, Gieles~(2009) developed a series of analytic solutions for the evolution of a Schechter mass function due to cluster disruption, and showed that this provided a good fit to the age/mass distributions of clusters in M51.

Whitmore et al.~(2010) did not see a truncation in the mass function of clusters in the Antennae galaxies based on power-law fits to the data, although we note that their last two data points lie below the predictions of a pure-power law.  A lack of a truncation was also reported in the cluster populations of M83 (C10) and M51 (Chandar et al.~2011).  

\begin{figure}
\includegraphics[width=8.5cm]{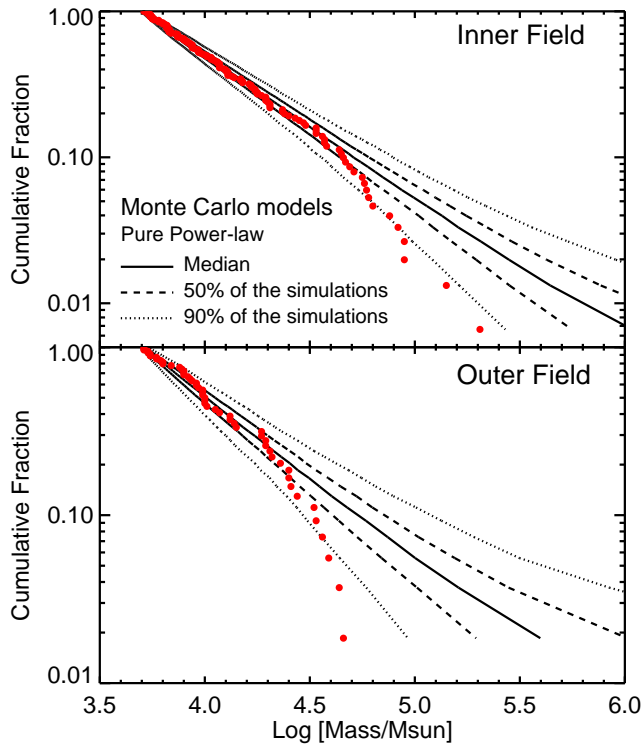}
\caption{The distribution of cluster masses with ages between 3 and 100~Myr.  Filled (red) circles represent the observations for each field.  The lines denote the results of a series of Monte Carlo simulations that adopt the same number of clusters as the observations (only those used in the figure) for a pure power-law mass distribution with an index of -2 and no upper mass truncation. The solid line shows the median mass expected, while the dashed and dotted lines enclose 50\% and 90\% of the simulations, respectively.  Note that both the inner and outer field distributions are inconsistent with the pure power-law case and require a truncation in the mass function at high masses.}
\label{fig:mass_func_pl}
\end{figure}

Since the truncation in the mass function that has been proposed previously is generally near the end of the distribution, low number statistics often can mask such a truncation, especially if the data is binned.  To avoid such an artefact from entering into our analysis, we will study the cumulative distribution of stellar clusters in M83.  The derived cluster mass distributions for the inner (top panel) and outer (bottom panel) fields of M83 for clusters with ages between 3 and 100~Myr are shown in Fig.~\ref{fig:mass_func_pl} as filled (red) circles.  If the distribution is described by a pure power-law (i.e. without truncation) the points should follow a straight line.  However, it is clear that both fields show a bend in their distributions.

In order to quantify the apparent bend, we have created a suite of Monte Carlo models.  We stochastically sample from a power-law mass function with an index of $-2$, sampling the same number of clusters as the observed distributions (i.e. the number of clusters with ages between 3 and 100~Myr and have masses above $5\times10^3$\msun).  Each field was treated separately.  The lines in Fig.~\ref{fig:mass_func_pl} show the results of the simulations, with the solid line representing the median of 2000 simulations, and the dashed and dotted lines enclosing 50\% and 90\% of the simulations, respectively.  The data appear to be inconsistent with the models at the $\gtrsim2\sigma$ level, suggesting a truncation at high masses is required.

\begin{figure}
\includegraphics[width=8.5cm]{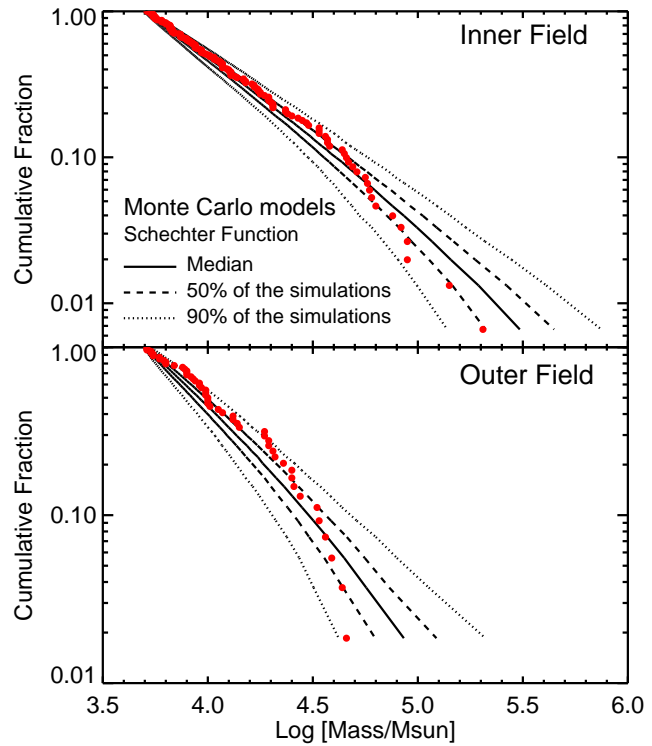}
\caption{The same as Fig.~\ref{fig:mass_func_pl}, but now the line represent Monte Carlo models sampling from a Schechter mass function. The simulations adopt truncation masses, M$_c$, of  $5\times10^5$ and $1\times10^5$\msun, respectively.}
\label{fig:mass_func_sch}
\end{figure}

In Fig.~\ref{fig:mass_func_sch} we show the results of a similar set of Monte Carlo simulations that stochastically sample from an underlying Schechter cluster mass function.  The simulations adopt a power-law with index of $-2$ on the low mass end and a truncation mass, M$_c$, of  $5\times10^5$ and $1\times10^5$\msun\ for the inner and outer fields, respectively.  In both sets of Monte Carlo models cluster disruption was not included, which may explain why the mass function does not fit the data of Field~2 for the Schechter function case.  This will be treated in more detail in \S~\ref{sec:disruption}.

Although disruption is not included in Fig.~\ref{fig:mass_func_sch}, we note that disruption is unlikely to cause the observed truncation.  This is because disruption would have to affect the high mass end more than the low mass end of the distribution, in disagreement with both of the main empirically derived disruption laws as well as theoretical expectations (see \S~\ref{sec:disruption}).

We conclude that the mass function (MF) of clusters in M83 is not well fit by a pure power-law, but instead requires a truncation at the high mass end.  The exact mass of the truncation is degenerate to some degree with the amount of cluster disruption, which will be quantified in \S~\ref{sec:disruption}.  The cause of the truncation is not clear from the present data, but it is likely related to the form of the GMC mass function, from which clusters form.  Rosolowsky et al.~(2007) have shown that the mass function of GMCs in M33 is truncated at high masses.  Such a truncation in the GMC MF would then relate directly (if a cluster forms from a fixed fraction of the GMC mass) or indirectly to a truncation in the cluster MF, which would be lower by a given factor (the fraction of mass of the GMC turned into stars within the cluster).

\section{Mass distribution and cluster disruption}
\label{sec:disruption}

%\subsection{Expectations from cluster disruption models}

As discussed in \S~1, two main empirically based scenarios of cluster disruption have been put forward in the literature.  These two scenarios differ in the roles that the environment and cluster mass play in the cluster disruption process.  In the mass {\it independent} disruption scenario (MID), cluster disruption is driven largely by internal processes, hence the environment or cluster mass play a limited, if existent, role (e.g. Fall et al.~2009).  In the mass {\it dependent} disruption scenario (MDD), the average lifetime of a cluster increases with increasing mass (i.e. more massive clusters live longer), and the lifetime decreases for increasing GMC density or stronger tidal fields (e.g. Lamers et al.~2005).

The dataset presented here is ideal for differentiating between these two scenarios due to the sampling of a relatively large range of galactocentric distances (i.e., ambient environments) and large sample of clusters to limit Poissonian noise.  Below we describe the models and then apply them to the observations to provide new constraints.

\subsection{Model Description}

We create artificial age and mass distributions of cluster populations using the two disruption models MID and MDD. First a value for the (galaxy wide) SFR  needs to be chosen. Then a {\it cluster formation efficiency} $\Gamma$ (Bastian~2008) needs to be assumed, such that the cluster formation rate is $\cfr =\Gamma \times \sfr$. We use a Schechter type cluster initial mass function (CIMF) 
\begin{equation}
\dndmidt = A\,\mi^{-2}\,\exp\left(\frac{\mi}{\mstar}\right).%,\hspace{1cm} \mlo\le\mi\le\mup.
\end{equation}
Here $\dndmidttext$ is the probability of forming a cluster with an initial mass between $\mi$ and $\mi+\dr \mi$ in an interval $\dr t$. The (exponential) truncation of the mass function occurs at $\mstar$.  % and $\mlo$ and $\mup$ are the lower and upper limit respectively (note that $\mup>\mstar$).
The constant $A$ relates to the $\cfr$ as $A=\cfr/\eone(\mlo/\mstar)$, where $\mlo$ is the lower limit of the mass function and $\en(x)$ with $n=1$ is a generalised expression of the exponential integral (Gieles~2009). 

The age dependent mass function at older ages can be described by so-called {\it evolved Schechter functions} (Jord\'{a}n et al.~2007). For this we need to find the evolution of the mass of a cluster as a function of the MDD and MID parameters.
For this we use the descriptions given in Gieles~(2009) and Larsen~(2009), respectively. We here introduce the evolved Schechter function for the combined evolution, including MDD and MID. 

The mass-loss rate of the cluster, $\mdot$, constitutes the mass loss rates due to MDD, $\mdotd$, and the one due to MID, $\mdoti$

\begin{equation}
\mdot=\mdotd+\mdoti.
\end{equation}
For $\mdotd$ we use the formalism of Lamers et al.~(2005) and Gieles~(2009), namely that the disruption timescale is
$\tdis =\tn M^\gamma$, where $\gamma\simeq0.65$ and $\tn$ is a constant that depends on the environment. From this we find $\mdotd=-M/\tdis=-M^{1-\gamma}/\tn$. The mass evolution in the mass-independent disruption case is $M(t)=\mi(t/\ts)^\lambda$, where $\ts$ is the time that mass-independent disruption starts and $\lambda \equiv \log_{10}(1-\fmid)$, where $\fmid$ is the fraction of the mass that is lost per age dex. From this we find $\mdoti=\lambda M/t$. We note that we have assumed that in the mass-independent disruption case each cluster loses a certain fraction of its mass, rather than a certain fraction of clusters being completely destroyed, while the others are not affected.  If the cluster mass function is a pure power-law of $-2$, there is no distinction between these two interpretations, as they both result in the same distributions.  However, if there are features in the cluster mass function, such as in a Schechter function, then the results may differ.  Particularly, if MID is applied to the $dM/dt$ distribution (i.e. all clusters lose 90\% of their mass), then the Schechter mass decreases with time.  On the other hand, if it is applied to the $dN/dt$ distribution (i.e. 90\% of all clusters are destroyed, irrespective of mass) then the Schechter mass remains constant.

Combining the mass loss from the mass dependent and mass independent disruption models results in the total mass-loss rate of:

\begin{equation}
\mdot=-\frac{M^{1-\gamma}}{\tn} + \lambda\frac{M}{t}.
\end{equation}
This non-separable first order differential equation can be solved after a variable substituting $y=M^\gamma$ and using an integrating factor and the result is

\begin{equation}
M = \left(\mi^\gamma\left[\frac{t}{\ts}\right]^{\lambda\gamma}-\frac{\gamma}{1-\lambda\gamma}\frac{t}{\tn}\right)^{1/\gamma}.
\label{eq:mt}
\end{equation}
For convenience we introduce the new variables \mbox{$\funcmid\equiv(t/\ts)^\lambda$} and $\dlg\equiv(\gamma/[1-\lambda\gamma])(t/\tn)$ such that $M=([\mi\funcmid]^\gamma-\dlg)^{1/\gamma}$. Note that for no MID, i.e. $\fmid=0$ we have $\funcmid=1$ and $\dlg=\gamma t/\tn$ and equation~(\ref{eq:mt}) reduces to the result of only MDD (equation 24 in Gieles~2009)

The final distribution of masses as a function of age can be found from conservation of number
\begin{equation}
\dndmdt=\dndmidt\dmdm,
\end{equation}

that is ,
\begin{equation}
\dndmdt=\frac{A\funcmid M^{\gamma-1}}{(M^\gamma+\dlg)^{1+1/\gamma}}\exp\left(-\frac{[M^\gamma+\dlg]^{1/\gamma}}{\funcmid\mstar}\right).
\end{equation}

To create the age distribution of a population we simply integrate over all masses, i.e.

\begin{equation}
\dndt=\int_{\mlo}^{\mup}\dndmdt dM,
\end{equation}
and equivalently for the mass distribution we find 

\begin{equation}
\dndm=\int_{\tmin}^{\tmax}\dndmdt dt.
\end{equation}

\subsection{Fitting the Observations}
\label{sec:fitting}

As in B11, we chose to split our catalogue into three age bins (age $\le 10$~Myr, $10 < {\rm age} \le 100$~Myr and $100 < {\rm age} \le 1000$~Myr) and look at the mass distribution of the clusters in each bin.  Note that the oldest bin is somewhat incomplete at low masses.  This is the same technique used to study disruption in the SMC (de Grijs \& Goodwin~2008), LMC (Chandar et al.~2010a) and M83 (Chandar et al.~2010b).  The benefit of using this technique is that it allows the completeness limit to be seen relatively clearly.  Additionally, the number of clusters per bin is divided by the linear age range covered to get a normalised mass function.  If the cluster formation rate of a galaxy has been constant, and disruption has not affected the population, then the three mass distributions should lie nearly on top of each other, modulo shot noise and stellar evolutionary effects.  

We adopt equal number bins in order to avoid small-N biases (e.g.~Ma{\'{\i}}z Apell{\'a}niz 
\& {\'U}beda~2005). We use 8 clusters per bin, except for the oldest ages in the outer field where we used 11 per bin due to the larger scatter.  The observed distributions are shown in Fig.~\ref{fig:dndmdt_mid}.  As noted previously, the youngest age bin ($\le10$~Myr)  may be under-represented due to the difficulty in distinguishing clusters from associations.  

In the following sections we compare the observations to a set of models visually in order to derive characteristic values.  In all cases the vertical shift is normalised to the intermediate age bin ($10 < $ age $\le 100$~Myr).

\subsubsection{Only allowing mass independent disruption}

\begin{figure}
\includegraphics[width=8.5cm]{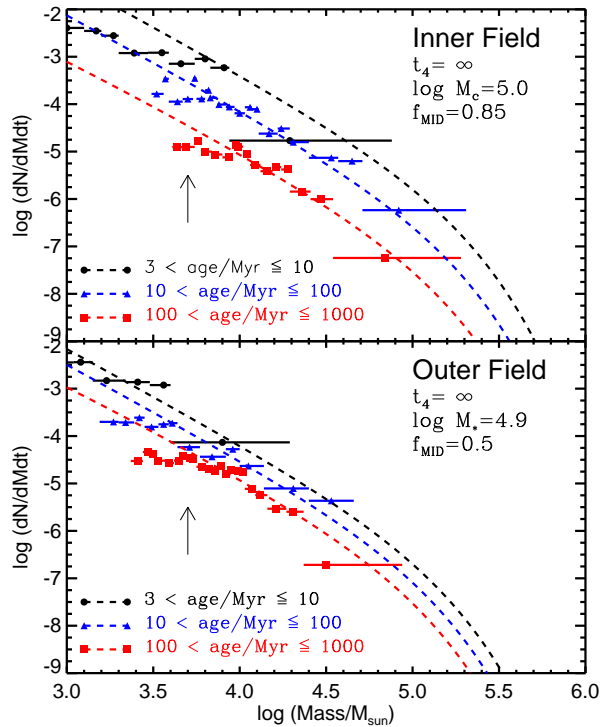}
\caption{The mass distributions for three age bins (normalised to the linear age range considered) for the inner (top panel) and outer (bottom panel) fields. The points represent the median mass in each mass bin, while the horizontal lines show the mass range covered by the bin. The dashed lines show the best fitting cluster population model assuming a Schechter mass distribution and only {\it mass independent} disruption (MID), normalised to the intermediate age bin.  We have used bins with equal numbers of clusters (8 per bin, except for the oldest ages in the outer field where we used 11 per bin due to larger scatter).  The arrow in each panel shows the lower mass limit of $5 \times10^3$\msun\ adopted.}
\label{fig:dndmdt_mid}
\end{figure}

First, we investigate models where only mass independent disruption is included.  The best fitting model is shown in Fig.~\ref{fig:dndmdt_mid}, along with the best fit parameters.  Our results for the inner field agree well with that found by C10, namely that the observed distributions can be well fit with purely mass independent disruption and that a Schechter function is not required to fit the data (although it also fits when a value of \mc$\sim10^5$~\msun\ is used).  However, as discussed in B11, the disruption rate in the outer field appears to be much lower.  A pure MID model does provide a good fit to the data, however instead of \fmid=$0.8-0.9$, a value of \fmid$=0.5$ is needed.  Even in the pure MID case, it is clear that environmentally dependent disruption is required to fit the data.  This is consistent with the Elmegreen \& Hunter~(2010) model of cluster disruption (see also Kruijssen et al.~2011a), which may be mass independent but is strongly dependent on the environment.

\subsubsection{Only allowing mass dependent disruption}

Next, we assume that cluster disruption is only dependent on cluster mass and that the initial distribution of cluster masses is described by a Schechter function with a characteristic mass, \mc\ (\S~\ref{sec:mass_func}).  As shown by Chandar et al.~(2010b), if the initial cluster mass function was a power-law with no truncation, and only mass dependent cluster disruption is taken into account, the three mass distributions (i.e. the 3 age bins) should converge at high masses.  Additionally, it is difficult to obtain the vertical offset between the intermediate and old age bins without invoking very short disruption timescales, which in turn is hard to reconcile with the relatively steep shape of the mass function in the oldest bins.

\begin{figure}
\includegraphics[width=8.5cm]{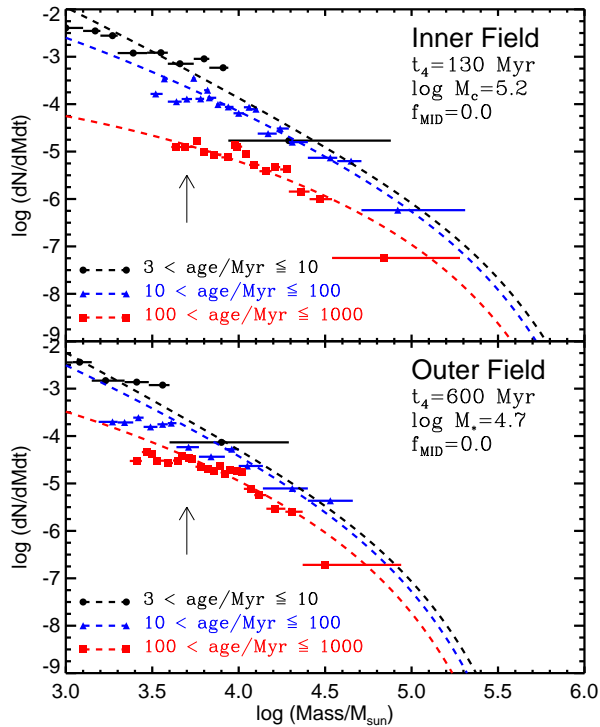}
\caption{The same as Fig.~\ref{fig:dndmdt_mid} except now the dashed lines show the best fit models for {\it mass dependent} (MDD) disruption.}
\label{fig:dndmdt_mdd}
\end{figure}

However, when a Schechter initial mass function is used, no such problems arise.  In Fig.~\ref{fig:dndmdt_mdd} we show the best fitting pure MDD model for the inner (top panel) and outer (bottom panel) fields.  The vertical arrow shows a mass of $5\times10^3$~\msun, our adopted lower mass limit.  We find that the observed distributions can be well fit with only mass dependent disruption and an initial Schechter mass function.  In general, for both fields, a value of \mc$\sim10^5$\msun\ is required to fit the data, and that the destruction rate of clusters is higher in the inner field than the outer field.  The derived characteristic timescale of cluster disruption is a factor of $\sim4$ times shorter in the inner field than the outer field.  This is in good agreement with numerical (Baumgardt \& Makino~2003) and analytic (Lamers, Baumgardt \& Gieles~2010) predictions, where the disruption timescale is proportional to the galactocentric distance, if tidal evaporation is mainly responsible for the dissolution.

\subsection{Maximum likelihood fitting}
\label{sec:max_like}

As noted above, however, simply comparing the models to the observations directly results in acceptable fits for both purely mass dependent and mass independent disruption.  Additionally, one can think of a hybrid model, where cluster disruption is mass independent for the first $\sim10$~Myr and afterwards becomes mass dependent (e.g.~Bastian et al.~2009a).  Due to the limited dynamic range of the data, it is presently not possible to definitively constrain the disruption law. However, the pure environmentally independent MID scenario that results in ``universal" age and mass distributions is disfavoured, given the observed differences between the two fields.  For the remainder of this section we focus on pure MDD models.

\begin{figure*}
\includegraphics[width=12cm]{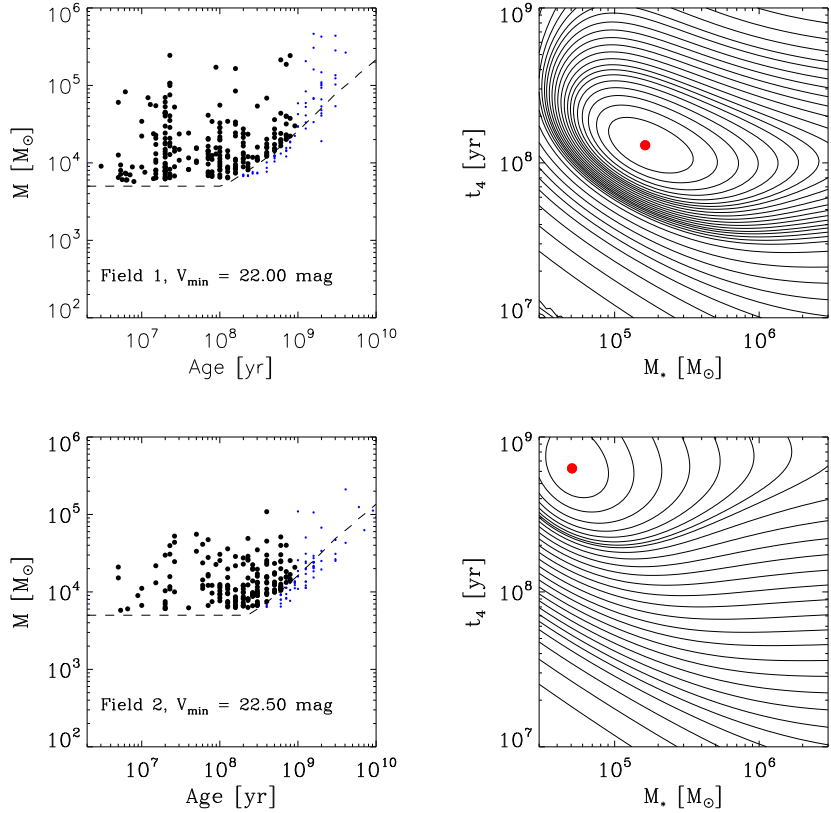}
\caption{Maximum likelihood fits for the inner (top) and outer (bottom) fields assuming purely mass dependent disruption. In the left panels, the adopted magnitude limit is shown, and the solid circles indicate clusters used in the fit (those with masses above $5\times10^3\msun$, have V-band magnitudes above the completeness limits, and have ages less than 1 Gyr), while the small points are clusters not included.  In the right panels, the best fitting model is shown with a filled (red) circle.  In both cases, a Schechter mass function is required to fit the data.}
\label{fig:maxlike}
\end{figure*}

In order to carry out a binning independent analysis, we have also compared the pure MDD models to the observations using a maximum likelihood fitting technique.  Here, we adopted $\gamma=0.65$ and fit on \t4\ and \mc.  The results are shown in Fig.~\ref{fig:maxlike}.  The best fitting model for the inner field has $\mc = 1.6 \times10^5$\msun and $\t4 = 130$ Myr, while for the outer field we find $\mc = 5 \times10^4$\msun and $\t4 = 600$ Myr.  The disruption timescale in the inner region of M83 is similar to that derived for the inner regions of the spiral galaxy, M51 (100-200~Myr; Gieles et al.~2005).

In both fields a Schechter function is required to fit the data, however \mc\ appears to be a factor of $\sim2-4$ larger in the inner field.  This suggests that the truncation mass may be dependent on environment, consistent with the discovery of significantly more massive clusters in starburst galaxies (e.g.~Maraston et al.~2004; Whitmore et al.~2010).  The characteristic mass derived for each field is consistent with that implied by the bend observed in the luminosity function discussed in \S~\ref{sec:lf}.

\begin{figure}
\includegraphics[width=8.5cm]{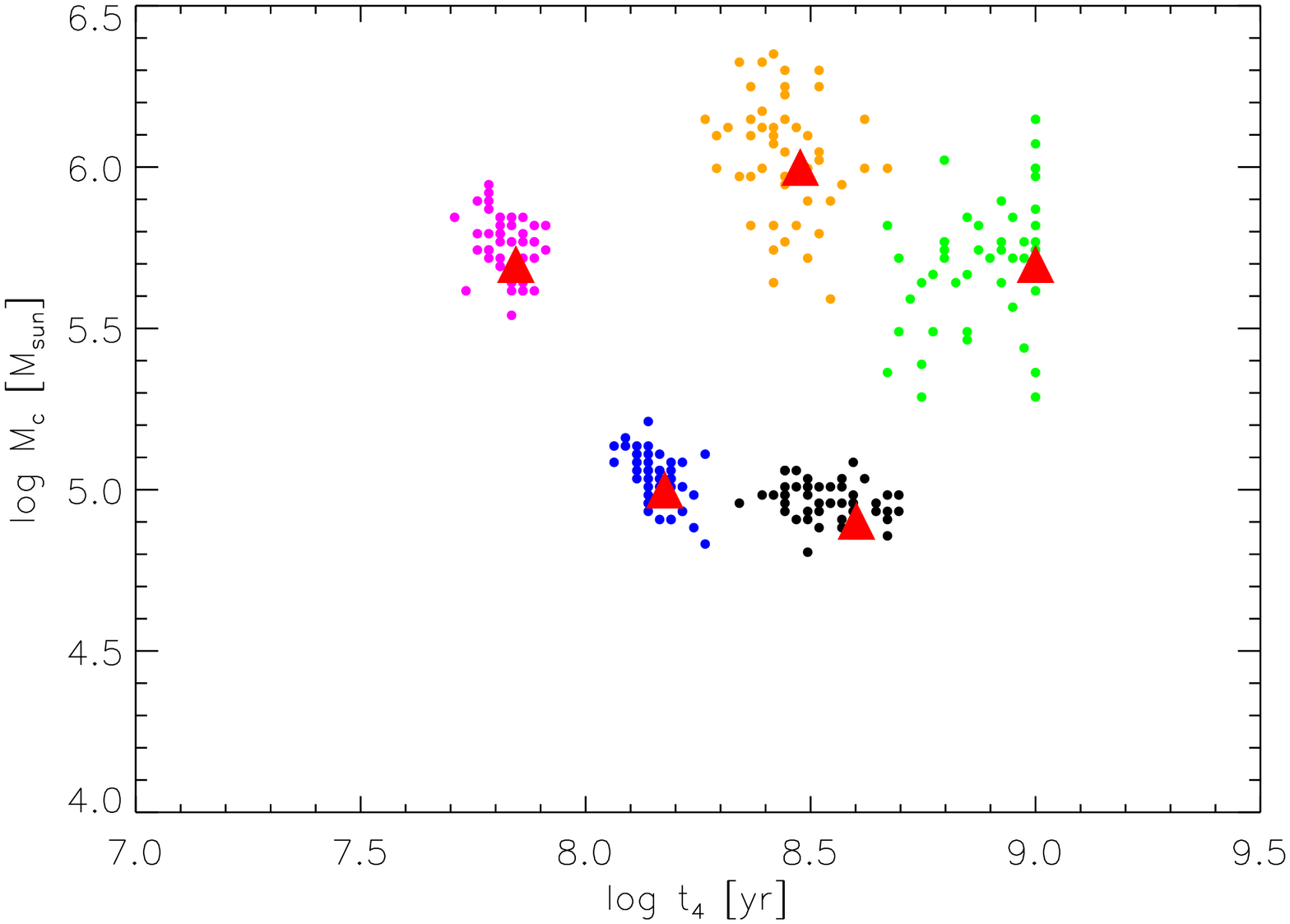}
\caption{The results of 50  Monte Carlo realisations of a cluster population that was stochastically generated with known parameters (\t4,\mc).  The same number of clusters as the observations were used, and were fit with the maximum likelihood technique.  The solid (red) triangles denote the input parameters while the small coloured circled denote the output values.  The method successfully recovers the input parameters.}
\label{fig:sim1}
\end{figure}

In order to test how robust our parameter determinations are, we have generated 50 realisations of stochastically generated cluster populations for five different combinations of \t4\ and \mc.  For each realisation we selected 380 clusters that match our selection criteria (i.e., are more massive than our lower mass limit) and use the same fitting code as was used on the observations.  The results are shown in Fig.~\ref{fig:sim1}.  We see that the fitting method can successfully recover the input values of \t4\ and \mc, increasing the level of confidence for a truncation in the MF, as well as for differing values of \t4\ and \mc\ for the two fields, although some small biases may remain.

%\begin{figure*}
%\includegraphics[width=10cm]{Figs/inner_contours_tot.ps}
%\caption{\rchi\ contours of the model fits to the observations discussed in \S~\ref{sec:full_comp}.  The lowest \rchi\ value is marked with a (red) asterix.  }
%\label{fig:inner_contours_tot}
%\end{figure*} 

%\begin{figure*}
%\includegraphics[width=10cm]{Figs/outer_contours_tot.ps}
%\caption{The same as Fig.~\ref{fig:inner_contours_tot} except now for the outer field.}
%\label{fig:outer_contours_tot}
%\end{figure*} 

%\clearpage

%\subsection{The lifetime and expansion of H$\alpha$ shells}

\section{Discussion: Infant mortality, clustered star formation, and implications for extragalactic surveys}
\label{sec:discussion}

%\subsection{Implications for Extragalactic Surveys}
%\label{sec:implications_exgal}

%why the Larsen relation (e.g. Bastian 2008) holds?  HST studies, focussed in on bright semi-resolved *clusters*

%\subsubsection{Infant mortality and the fraction of star-formation in clusters}
%\label{sec:im}

As noted throughout this work, obtaining a full census of young clusters ($<10$~Myr) in extragalactic studies is not possible, due to the difficulty in distinguishing clusters from unbound associations.  This echoes the results of Bressert et al.~(2010) and GPZ11 who showed that even within the Galaxy, it is not possible to distinguish clusters/associations until they are dynamically evolved.  This makes it difficult to directly estimate the fraction of stars that form that will end up as part of a bound cluster ($\Gamma$; Bastian~2008), hence indirect tracers may be required.  For extragalactic surveys, depending on the selection criteria adopted, the fraction of young stars in clusters can vary by a factor of two or more.  In Fig.~\ref{fig:example_chandar} a selection of young groups are shown that are not included in the cluster sample in the current work.  For each of these regions, it is possible (even likely) that some fraction of the young stars will end up in clusters, however it is not possible to estimate this fraction from the observations.  This effect may explain why estimates of $\Gamma$ of distant starburst galaxies are higher than seen in local galaxies (e.g. The Antennae - Fall et al.~2005; NGC~3256 - Goddard et al.~2010), although a physical difference between starbursts and quiescent galaxies is not ruled out (Adamo et al.~2011).

Including unbound associations (which by definition do not appear in older age bins) in the youngest age bins of cluster populations will lead to the interpretation that some fraction of the young ``clusters" must disrupt on short timescales.  This has contributed to the idea of infant mortality, and likely accounts for a significant portion of the observed drop between the youngest ($<10$~Myr) age bin and the following bins (e.g.~Fall et al.~2005; Bastian et al.~2005; Bonatto \& Bica~2011; Adamo et al.~2011).  If restrictive definitions of samples are used, the degree of infant mortality needed to explain a population is much lower than if loose definitions are used.

Since cluster formation is a highly dynamic process, and it is likely that clusters lose mass due to internal and external effects during their early evolution, the fraction of stars found in clusters may be strongly dependent on the age of the cluster population under study.  In the limiting case where clusters do not lose mass after their formation, one simply needs to look at evolved clusters.  Alternatively, if clusters lose mass or are destroyed rapidly during their early dynamical evolution, the fraction of stars in clusters may be impossible to define.

One way to distinguish between classical infant mortality and the inclusion of associations in the sample, is through measurements of the dynamical stability of young massive clusters (YMCs).  If infant mortality is driven by gas expulsion and it significantly affects dense massive clusters, then many/most YMCs should show large velocity dispersions due to their expansion.  Extragalactic YMCs indeed appear to be ``super-virial", which has been interpreted as evidence of their impending dissolution (Goodwin \& Bastian~2006).  However, Gieles, Sana \& Portegies Zwart~(2010) have shown that high-mass binaries are likely significantly contributing to the observed high velocity dispersions.  For YMCs within the Galaxy or the Magellanic Clouds, however, individual stars can be resolved for multi-epoch velocity monitoring (or proper motion studies) in order to remove binaries.  The sample is still small, but preliminary works on NGC~3603 (Rochau et al.~2010), the Arches (Clarkson et al.~2011; Clarkson et al. in prep), Westerlund 1 (Mengel \& Tacconi-Garman~2007; Cottaar et al. in prep), and R136 (H\'{e}nault-Brunet et al. in prep) indicate that they have low velocity dispersions, either in a virial or sub-virial state.  This suggests that they are already in dynamical equilibrium at a very young age ($<3$~Myr), consistent with the cluster formation simulations of Kruijssen et al.~(2011b), but in contrast with the infant mortality scenario (see the recent review by Bastian~(2011), who discuss this in more detail).  Larger samples of young clusters will be needed to confirm or refute the infant mortality scenario.

\section{Conclusions}
\label{sec:conclusions}

We have presented an analysis of the cluster population in M83 based on HST/WFC3 observations.  In particular, we have based our analysis largely on a comparison between the cluster populations in the inner and outer regions of the galaxies.  Our main conclusions are summarised below.

{\bf Sample selection:} We have taken a relatively conservative approach in identifying clusters, only using those that are resolved, centrally concentrated and symmetric. This removes many associations from the sample, but at the expense of missing some clusters that are within larger associations.  As shown by Gieles \& Portegies Zwart~(2011) it is not possible to differentiate between clusters and associations until they are dynamically evolved.  Additionally, when constructing extragalactic samples, it is often impossible to define the limits of clusters, or whether some parts of larger associations will become/remain bound.  As such, the definition of clusters is often ambiguous at young ages
 and it becomes difficult to construct unique samples, which likely causes much of the disagreement in the literature.  However, at ages older than $\sim10$~Myr, most clusters appear to be distinct objects, allowing for complete and unambiguous samples to be drawn.

{\bf Comparison with previous results:}  The cluster population in the inner field of the galaxy has been studied in detail in C10.  While the cluster samples used in C10 and in the present work differ at the youngest ages (see \S~\ref{sec:comparison}), we find good agreement between our results and theirs after $\sim10$~Myr.  Before this age, the samples disagree due to 1) selection criteria and 2) the inclusion of the nuclear region of the galaxy in C10.  After $\sim10$~Myr, the age/mass distributions derived for the inner field in the present work are consistent with those of C10.

{\bf $\Pi$ distribution:}  We have compared our cluster sample with the dynamical definition of a cluster given by Gieles \& Portegies Zwart~(2011), namely that $\Pi$ ($t_{\rm age}/t_{\rm cross}$) is greater than 1.  We find that the vast majority of the objects ($>95$\%) that were selected through visual inspection are clusters according to the $\Pi$ definition.  The fraction of clusters that were missed is not possible to estimate at the moment, since it is difficult to define a characteristic size within crowded regions or associations.

{\bf Size distribution:}  We find that the size distribution of clusters in M83 is adequately fit by a log-normal distribution with a peak at 2.3~pc and 2.6~pc in the inner and outer fields, respectively.  However, below $\sim0.5$~pc our catalogue may be incomplete due to the inability to resolve clusters from point sources.  Additionally, we find a trend between age and cluster size, such that older clusters are larger on average.  Whether this is a physical effect or due to the fitting procedure is uncertain.

{\bf Luminosity distribution:}  The luminosity distribution of clusters in the U, B, V and I bands can be approximated by a power-law distribution with index, $-\alpha$.  Over the full range where we are complete, we find $1.8 \leq \alpha \leq 2.1$, similar to that seen in other studies, with the bluer bands having shallower indices.  Additionally, we find evidence of a truncation or bend in the luminosity function at $M_V = -9.3$ and $-8.8$ for the inner and outer fields, respectively.  This is similar to that seen in the cluster populations of M51 and NGC~6946.

{\bf Mass distribution:}  The distribution of cluster masses also appears to follow a power-law with an index of $\sim -2$ over most of the observed range.  However, the upper end of the mass function appears to be truncated.  This is shown using cumulative fractions, however we note that the effect is not seen clearly if binning is used.  Monte carlo simulations are used to show the significance of the results.  Including disruption, we find best fit values of $1.6 \times 10^5$\msun\ and $5\times10^4$\msun\ for the inner and outer fields, respectively.  These values are consistent with that implied by the observed bends in the luminosity distributions.  This suggests that the truncation mass is dependent on environment, likely related to the maximum mass of GMCs.

{\bf Cluster disruption:}  Finally, we have studied the role of cluster disruption in shaping the population.  We have tested the observations against predictions from {\it mass dependent disruption} models (MDD) as well as {\it mass independent disruption} (MID).  We have also presented a formulation of how the two models can be combined.  We find that both the MDD and MID models can provide adequate fits to the data, if a Schechter mass function is used.  If the pure MID model is used, the fraction of clusters disrupted every decade in age varies as a function of environment, in apparent contradiction to the ``universal" age and mass distributions predicted by the preferred form of MID.  If the pure MDD model is used, we find that the characteristic disruption timescale is a factor of $\sim4$ longer in the outer field, in agreement with analytic and numerical predictions.

\section*{Acknowledgments}

We thank Diederik Kruijssen for insightful discussions.  Additionally, we would like to thank Rupali Chandar and Brad Whitmore for providing access to their catalogues of cluster candidates in M83 as well as helpful discussions.  MG acknowledges financial support from the Royal Society.

\bsp
\label{lastpage}
\end{document}